\documentclass[aps,showpacs,preprint,footinbib,preprintnumbers]{revtex4}

\def\ltap{\ \raise.3ex\hbox{$<$\kern-.75em\lower1ex\hbox{$\sim$}}\ }
\def\gtap{\ \raise.3ex\hbox{$>$\kern-.75em\lower1ex\hbox{$\sim$}}\ }
\def\gl{\ \raise.4ex\hbox{$>$\kern-.75em\lower1ex\hbox{$<$}}\ }

\usepackage[normalem]{ulem}  
\usepackage[dvips]{color} 
\usepackage{amsmath,amssymb}
\usepackage{booktabs}
\usepackage{mathrsfs}
\usepackage{graphicx}
\usepackage{multirow}

\renewcommand\sout{\bgroup \color{red} \ULdepth=-.5ex \ULset}

\newcommand{\Ex}[2]{\ifmmode{#1\times10^{#2}}\else{$#1\times10^{#2}$}\fi}

\begin{document}

\title{Exotic mesons with double charm and bottom flavor}
\author{S.~Ohkoda$^1$, Y.~Yamaguchi$^1$, S.~Yasui$^2$, K.~Sudoh$^3$ and A.~Hosaka$^1$}
\affiliation{$^1$Research Center for Nuclear Physics (RCNP), 
Osaka University, Ibaraki, Osaka, 567-0047, Japan}
\affiliation{$^2$KEK Theory Center, Institute of Particle and Nuclear
Studies, High Energy Accelerator Research Organization, 1-1, Oho,
Ibaraki, 305-0801, Japan}
\affiliation{$^3$Nishogakusha University, 6-16, Sanbancho, Chiyoda,
Tokyo, 102-8336, Japan}
\date{\today}

\begin{abstract}
We study exotic mesons with double charm and bottom flavor
 ($|C|=|B|=2$), whose quark configuration is $\bar{\mathrm
 Q}\bar{\mathrm Q}{\mathrm q}{\mathrm q}$. This quark configuration has
 no annihilation process of quark and antiquark, and hence is a  genuinely exotic states.
We take a hadronic picture by considering the molecular states composed of
 a pair of heavy mesons, such as $\mathrm{D}\mathrm{D}$,
 $\mathrm{D}\mathrm{D}^*$ and $\mathrm{D}^*\mathrm{D}^*$ for charm flavor, and $\mathrm{B}\mathrm{B}$,
 $\mathrm{B}\mathrm{B}^*$ and $\mathrm{B}^*\mathrm{B}^*$ for bottom flavor.
The interactions between heavy mesons are derived
 from the heavy quark effective theory.
All molecular states are classified by $I(J^P)$ quantum numbers, and are systematically studied up to the total angular momentum $J \leq 2$.
By solving the coupled channel Schr\"odinger equations,
due to the strong tensor force of one pion exchanging, we find  bound
 and/or resonant states of various quantum numbers.
\end{abstract}
\maketitle

\section{Introduction}

Exotic hadrons which include multiquark configurations provide us with important information in hadron physics.
There is a key to understand one of the most important problems in
 hadron physics; what are the constituent particles of hadrons, and what
are the interactions among the constituent particles at relevant low energies.
Those questions are related to the fundamental questions of the QCD,
such as to color confinement, chiral symmetry breaking and so on.
Nowadays the exotic hadrons are studied, not only in light flavor
sectors, but also in heavy flavor sectors with charm and bottom quarks \cite{Brambilla:2004wf,Swanson:2006st,Voloshin:2007dx,Nielsen:2009uh,Brambilla:2010cs}.
Recent experimental observations of heavy exotic hadrons,
$\mathrm{D}_{\mathrm{sJ}}$, $\mathrm{X}$, $\mathrm{Y}$,
$\mathrm{Z}^{\pm}$ in charm sector, and $\mathrm{Y}_{\mathrm{b}}$,
$\mathrm{Z}_{\mathrm{b}}^{\pm}$ in bottom sector have motivated
intensive discussions about  possible new dynamics in the heavy hadrons.
Many of those hadrons have unusual mass, decay width and branching ratios, which  may not be explained as normal hadrons, such as $\bar{\mathrm q}\mathrm{q}$ and $\mathrm{q}\mathrm{q}\mathrm{q}$.
As candidates of flavor exotics for future experiments, a new hadron state $\mathrm{T}_{\mathrm{Q}\mathrm{Q}}$ whose quark content is $\bar{\mathrm{Q}}\bar{\mathrm{Q}}\mathrm{q}\mathrm{q}$ has been discussed theoretically \cite{Zouzou:1986qh,Lipkin:1986dw,Heller:1986bt,Carlson:1987hh,SilvestreBrac:1993ss,Semay:1994ht,Stanku97,Stanku98,Silvestre-Brac:1993ry,Stancu96,SchaffnerBielich:1998ci,Janc04,Barnea:2006sd,Vijande:2007fc,Vijande:2007rf,Ebert:2007rn,Navarra:2007yw,Lee:2007tn,Lee:2009rt,Zhang:2007mu,Yang:2009zzp,Vijande:2009kj,Carames:2011zz,Vijande:2011zz,Ader:1981db,Vijande:2007ix,Manohar:1992nd,Tornqvist:1993ng,Ding:2009vj,Molina:2010tx}.
$\mathrm{T}_{\mathrm{Q}\mathrm{Q}}$ is a system containing two heavy
quarks and it is genuinely a flavor exotic which cannot be assigned by a normal hadron.
In the present paper, we discuss the energy spectrum of the possible bound and/or resonant states of $\mathrm{T}_{\mathrm{Q}\mathrm{Q}}$.

In phenomenological studies, there exist two approaches to $\mathrm{T}_{\mathrm{Q}\mathrm{Q}}$ state.
In one approach,
 $\mathrm{T}_{\mathrm{Q}\mathrm{Q}}$ is considered as a
tetraquark state, in which the effective degrees of freedom are constituent quarks \cite{Zouzou:1986qh,Lipkin:1986dw,Heller:1986bt,Carlson:1987hh,SilvestreBrac:1993ss,Semay:1994ht,Stanku97,Stanku98,Silvestre-Brac:1993ry,Stancu96,SchaffnerBielich:1998ci,Janc04,Barnea:2006sd,Vijande:2007fc,Vijande:2007rf,Ebert:2007rn,Navarra:2007yw,Lee:2007tn,Lee:2009rt,Zhang:2007mu,Yang:2009zzp,Vijande:2009kj,Carames:2011zz,Vijande:2011zz,Ader:1981db,Vijande:2007ix}.
It is shown that $\mathrm{T}_{\mathrm{Q}\mathrm{Q}}$ may be a stable object due to the strong attraction in $\mathrm{q}\mathrm{q}$ which may form a
stable scalar diquark \cite{Jaffe:1976ig,Jaffe:1977cv,Jaffe03}.
As a result, $\mathrm{T}_{\mathrm{Q}\mathrm{Q}}$ may be a deeply bound state,
which does not decay through the  strong interaction \cite{Navarra:2007yw,Lee:2007tn,Lee:2009rt}.
If the diquark developed, the study of $\mathrm{T}_{\mathrm{Q}\mathrm{Q}}$ is also useful to understand the color superconductivity in high density quark matter \cite{Alford:1997zt,Rapp:1997zu,Alford:2007xm}.
The tetraquark states such as
$\mathrm{T}^{1}_{\mathrm{c}\mathrm{c}}$,
$\mathrm{T}^{1}_{\mathrm{c}\mathrm{b}}$ and
$\mathrm{T}^{1}_{\mathrm{b}\mathrm{b}}$ states with $I(J^P)=0(1^{+})$ as
well as  $\mathrm{T}^{0}_{\mathrm{c}\mathrm{b}}$  with $I(J^P)=0(0^+)$
have been discussed also as stable objects \cite{Lee:2007tn,Lee:2009rt}.

Another approach is the hadronic molecule picture.
When four quarks ($\bar{\mathrm{Q}}\bar{\mathrm{Q}}\mathrm{q}\mathrm{q}$) are present, they may form 
hadronic clusters ($\bar{\mathrm{Q}}\mathrm{q}$) which may alternatively become relevant degrees of freedom
instead of diquarks \cite{Manohar:1992nd,Tornqvist:1993ng,Ding:2009vj,Molina:2010tx}.
The hadronic molecule picture is applied to the energy region close to the thresholds.
In the $\mathrm{T}_{\mathrm{QQ}}$ system, two mesons composed by $\bar{\mathrm{Q}}\mathrm{q}$, the
pseudoscalar meson $\mathrm{P} \sim
(\bar{\mathrm{Q}}\mathrm{q})_{\mathrm{spin}\, 0}$ and the vector meson
$\mathrm{P}^{\ast} \sim (\bar{\mathrm{Q}}\mathrm{q})_{\mathrm{spin}\,
1}$, can become effective degrees of freedom as constituents.
In the heavy quark limit, the pseudoscalar meson $\mathrm{P}$ and the vector meson $\mathrm{P}^{\ast}$ become degenerate in mass, and hence both of them should be considered.
Hereafter we introduce the notation $\mathrm{P}^{(\ast)}$ for $\mathrm{P}$ and $\mathrm{P}^{\ast}$.
Indeed, we will show that the mass degeneracy of the $\mathrm{P}$ and $\mathrm{P}^{\ast}$ activates the one-pion exchange potential between two $\mathrm{P}^{(\ast)}$'s, and the bound and/or resonant $\mathrm{P}^{(\ast)}\mathrm{P}^{(\ast)}$ states are formed.

The tetraquark picture and the hadronic molecule picture are quite different.
The tetraquark picture is applied to the deeply bound state.
To apply this picture to the shallow bound states or resonant states, slightly below or above thresholds respectively, is not appropriate, because the continuous $\mathrm{P}^{(\ast)}$ states above thresholds are not taken into account.
On the other hand, the hadronic molecule picture can be applied to the shallow and resonant states, while it cannot be applied to the deeply bound states.
 

Though there have been several studies for bound states of $\mathrm{P}^{(\ast)}\mathrm{P}^{(\ast)}$ in the hadronic molecule picture,
resonant states of $\mathrm{P}^{(\ast)}\mathrm{P}^{(\ast)}$ have not been studied yet.
Moreover, the quantum numbers $I(J^{P})$ which have been discussed are limited only $J \le 1$.
To be more problematic, several channels in coupled channel problem have been neglected.
The last point is very important in the heavy quark systems.
In the heavy quark limit, we need to consider the mass degeneracy of $\mathrm{P}$ and $\mathrm{P}^{\ast}$ which provides more number of channels than discussed in the literature.
In the present work, we discuss both bound and resonant states of $\mathrm{P}^{(\ast)}\mathrm{P}^{(\ast)}$ systematically for various quantum numbers $I(J^{P})$ up to $J \le 2$ by considering the fully coupled channel problem.
The systematic analysis of the energy spectrum for various quantum numbers is important to investigate the dynamics governing the systems.
Indeed, it will be shown that there are several new shallow bound and/or resonant states, which were not found in other studies, thanks to the strong attraction induced from the channel couplings even for larger $J$ or angular momentum.

This paper is organized as follows.
In section 2, we give the interaction between two $\mathrm{P}^{(\ast)}$ mesons with respecting the heavy quark symmetry and chiral symmetry.
We introduce the two types of potentials, the $\pi$ exchange potential and $\pi \rho \, \omega$ exchange potential.
In section 3, we classify all the $\mathrm{P}^{(\ast)}\mathrm{P}^{(\ast)}$ systems up to $J=2$, and search the bound and/or resonant states by applying the potentials and solving the Schr\"odinger equations numerically.
In section 4, we compare our results from the hadronic molecule picture with the previous results from the tetraquark picture.
We summarize our discussions in the final section.

\section{Interaction with two mesons with doubly heavy flavor}

The dynamics of the hadronic molecule of $\mathrm{P}^{(\ast)}\mathrm{P}^{(\ast)}$ respects two important symmetries; the heavy quark symmetry and chiral symmetry.
The heavy quark symmetry induces the mass degeneracy of $\mathrm{P}$ and $\mathrm{P}^{\ast}$ in the heavy quark limit.
Because of this, we have to consider the channels of
degenerate pairs, such as $\mathrm{P}\mathrm{P}$, $\mathrm{P}\mathrm{P}^{\ast}$, $\mathrm{P}^{\ast}\mathrm{P}$ and $\mathrm{P}^{\ast}\mathrm{P}^{\ast}$, leading to the mixing of them; $\mathrm{P}\mathrm{P}^{\ast}$-$\mathrm{P}^{\ast}\mathrm{P}$, $\mathrm{P}^{\ast}\mathrm{P}^{\ast}$-$\mathrm{P}^{\ast}\mathrm{P}^{\ast}$, $\mathrm{P}\mathrm{P}$-$\mathrm{P}^{\ast}\mathrm{P}^{\ast}$, $\mathrm{P}\mathrm{P}^{\ast}$-$\mathrm{P}^{\ast}\mathrm{P}^{\ast}$.

As for the meson-exchange interaction between two $\mathrm{P}^{(\ast)}$'s, the one pion
exchange potential (OPEP) exists at long distances.
The existence of a pion is a robust consequence of spontaneous breaking of chiral symmetry \cite{Nambu:1961tp}.
The OPEP is provided by the $\mathrm{P}\mathrm{P}^{\ast}\pi$ and
$\mathrm{P}^{\ast}\mathrm{P}^{\ast}\pi$ vertices whose coupling strengths  are equally weighted thanks to the heavy quark symmetry.
We note that there is no $\mathrm{P}\mathrm{P}\pi$ vertex because of the parity conservation.
We should keep in mind in the following discussions that the existence of both $\mathrm{P}\mathrm{P}^{\ast}\pi$ and
$\mathrm{P}^{\ast}\mathrm{P}^{\ast}\pi$ vertices thanks to the heavy quark symmetry provide the channel mixings in $\mathrm{P}\mathrm{P}$, $\mathrm{P}\mathrm{P}^{\ast}$, $\mathrm{P}^{\ast}\mathrm{P}$ and $\mathrm{P}^{\ast}\mathrm{P}^{\ast}$ at long distance.
At short distances, heavier mesons exchange potential, which provide similar channel mixings, should also be considered.

To derive the $\mathrm{P}^{(\ast)}\mathrm{P}^{(\ast)}$ potential, 
we employ the effective Lagrangians based on the heavy quark
symmetry and  chiral symmetry~\cite{Burdman:1992gh,Wise:1992hn,Yan:1992gz,Casalbuoni:1996pg,Manohar:2000dt,Isola:2003fh}.
They describe the interaction between heavy mesons which are given
by  the exchange of pion and vector
mesons ($v = \rho ,\omega$). The interaction Lagrangians are given as
\begin{eqnarray}
 {\cal L}_{\pi PP^*} &=& 
 2 \frac{g}{f_\pi}(P^\dagger_a P^\ast_{b\,\mu}+P^{\ast\,\dagger}_{a\,\mu}P_b)\partial^\mu\hat{\pi}_{ab}
 \, , 
 \label{eq:piPP*}\\
 {\cal L}_{\pi P^*P^*} &=& 
 2 i \frac{g}{f_\pi}\epsilon^{\alpha \beta \mu \nu} v_{\alpha}
 P^{\ast\,\dagger}_{a\,\beta}P^\ast_{b\,\mu}\partial_{\nu}
 \hat{\pi}_{ab}
\label{eq:piP*P*} \, , \\
{\cal L}_{vPP} &=&
-\sqrt{2}\beta g_V P_b P^{\dagger}_a v \cdot \hat{\rho}_{ba}
 \, , 
\label{eq:vPP} \\
{\cal L}_{vPP^*} &=& 
-2\sqrt{2}\lambda g_V v_{\mu}\epsilon^{\mu \nu \alpha \beta}
\left(P^\dagger_a
P^\ast_{b\,\beta}-P^{\ast\,\dagger}_{a\,\beta}P_b\right)
\partial_\nu(\hat{\rho}_\alpha)_{ba}
\, ,
\label{eq:vPP*} \\
{\cal L}_{vP^*P^*} &=&
\sqrt{2} \beta g_V P^*_b P^{*\dagger}_a v \cdot \hat{\rho}_{ba}\nonumber \\
&&+i2 \sqrt{2}\lambda g_V
P^{\ast\,\dagger}_{a\,\mu}P^\ast_{b\,\nu}
(\partial^\mu(\hat{\rho}^\nu)_{ba}-\partial^\nu(\hat{\rho}^\mu)_{ba}) \, ,
\label{eq:vP*P*}
\end{eqnarray}
where $P= D$, $B$ and $P^*_{\mu} = D^*_{\mu}$, $B^*_{\mu}$.
The subscripts $a$ and $b$ are for light flavor indices, up and down, and $v_{\mu}$ is a four-velocity which will be fixed as $v_{\mu}=(1,\vec{0})$ below.
The pion and vector meson ($\rho$ and $\omega$) fields are defined by
\begin{align}
&\hat{\pi} = \left(
\begin{array}{cc}
 \frac{\pi^{0}}{\sqrt{2}} & \pi^{+} \\
 \pi^{-} & -\frac{\pi^{0}}{\sqrt{2}}
\end{array}
\right),
 &  \hat{\rho}_{\mu} =  \displaystyle{\left(
\begin{array}{cc}
 \frac{\rho^{0}}{\sqrt{2}} + \frac{\omega}{\sqrt{2}}& \rho^{+} \\
 \rho^{-} & -\frac{\rho^{0}}{\sqrt{2}} + \frac{\omega}{\sqrt{2} }
\end{array}
\right)}_{\mu} \, .
\end{align}
Following Ref.~\cite{Isola:2003fh}, the coupling constants in the
interaction Lagrangians are given as
\begin{align}
  g &= 0.59, & \beta &= 0.9, & \lambda &= 0.56 \,\mathrm{GeV}^{-1},
  & f_{\pi} &= 132 \, \mathrm{MeV}, & g_V &= \frac{m_{\rho}}{f_{\pi}},
\end{align}
where $f_{\pi}$ is the pion decay constant and $m_{\rho}$ is the mass of
the $\rho$ meson.
The OPEPs  are derived by the interaction Lagrangians
(\ref{eq:piPP*}) and (\ref{eq:piP*P*}) as follows:
\begin{eqnarray}
V^{\pi}_{P_{1}P_{2}^{\ast} \rightarrow P_{1}^{\ast}P_{2}} &\!=\!&
 \left( \sqrt{2} \frac{g}{f_{\pi}} \right)^{2} \frac{1}{3} \left[ \vec{\varepsilon}_{1}^{\,\ast} \!\cdot\! \vec{\varepsilon}_{2} \, C(r;m_{\pi}) \!+\! S_{\varepsilon_{1}^{\ast},\varepsilon_{2}} \, T(r;m_{\pi}) \right]  \vec{\tau}_{1}  \!\cdot\! \vec{\tau}_{2}, \label{eq:pot_PP*P*P}  \\
V^{\pi}_{P_{1}^{\ast}P_{2}^{\ast} \rightarrow P_{1}^{\ast}P_{2}^{\ast}} &\!=\!&
 \left( \sqrt{2} \frac{g}{f_{\pi}} \right)^{2} \frac{1}{3} \left[
   \vec{T}_{1} \!\cdot\! \vec{T}_{2} \, C(r;m_{\pi}) \!+\!
   S_{T_{1},T_{2}} \, T(r;m_{\pi}) \right]  \vec{\tau}_{1}  \!\cdot\!
 \vec{\tau}_{2}, \label{eq:pot_P*P*P*P*} \\
V^{\pi}_{P_{1}P_{2} \rightarrow P_{1}^{\ast}P_{2}^{\ast}} &\!=\!&
 \left( \sqrt{2} \frac{g}{f_{\pi}} \right)^{2} \frac{1}{3} \left[ \vec{\varepsilon}_{1}^{\,\ast} \!\cdot\! \vec{\varepsilon}_{2}^{\, \ast} \, C(r;m_{\pi}) \!+\! S_{\varepsilon_{1}^{\ast},\varepsilon_{2}^{\ast}} \, T(r;m_{\pi}) \right] \vec{\tau}_{1}  \!\cdot\! \vec{\tau}_{2}, \label{eq:pot_PPP*P*} \\
V^{\pi}_{P_{1}P_{2}^{\ast} \rightarrow P_{1}^{\ast}P_{2}^{\ast}} &\!=\!&
 -\left( \sqrt{2} \frac{g}{f_{\pi}} \right)^{2} \frac{1}{3} \left[
  \vec{\varepsilon}_{1}^{\,\ast} \!\cdot\! \vec{T}_{2} \, C(r;m_{\pi})
  \!+\! S_{\varepsilon_{1}^{\ast},T_{2}} \, T(r;m_{\pi}) \right]
 \vec{\tau}_{1}  \!\cdot\! \vec{\tau}_{2}, \label{eq:pot_PP*P*P*}
\end{eqnarray}
where $m_{\pi}$ is the pion mass.
Here three polarizations are introduced for $\mathrm{P}^{*}$ as defined by
$\vec{\varepsilon}^{\hspace{0.2em}(\pm)} \!=\! \left(\mp 1/\sqrt{2}, - i/\sqrt{2}, 0 \right)$ and
$\vec{\varepsilon}^{\hspace{0.2em}(0)} \!=\! \left(0, 0, 1\right)$,
and the spin-one operator $\vec{T}$ is defined by $T_{\lambda' \lambda}^{i}=i \varepsilon^{ijk} \varepsilon_{j}^{(\lambda')\dag} \varepsilon_{k}^{(\lambda)}$.
As a convention, we assign $\vec{\varepsilon}^{\,(\lambda)}$ for an
incoming vector particle and $\vec{\varepsilon}^{\,(\lambda)\ast}$ for
an outgoing vector particle.
Here $\vec{\tau}_{1}$ and $\vec{\tau}_{2}$ are isospin operators for
$\mathrm{P}^{(\ast)}_{1}$ and $\mathrm{P}^{(\ast)}_{2}$; $\vec{\tau}_{1}  \!\cdot\!
\vec{\tau}_{2} = -3$ and 1 for the $I=0$ and $I=1$ channels, respectively.  We define the tensor operators 
\begin{eqnarray}
S_{\varepsilon_{1}^{\ast},\varepsilon_{2}} &=& 3 ( \vec{\varepsilon}^{\,(\lambda_{1})\ast} \!\cdot\!\hat{r} ) ( \vec{\varepsilon}^{\,(\lambda_{2})} \!\cdot\!\hat{r} ) -  \vec{\varepsilon}^{\,(\lambda_{1})\ast} \!\cdot\! \vec{\varepsilon}^{\,(\lambda_{2})}, \\
S_{T_{1},T_{2}} &=& 3 ( \vec{T}_{1} \!\cdot\!\hat{r} ) ( \vec{T}_{2} \!\cdot\!\hat{r} ) - \vec{T}_{1} \!\cdot\! \vec{T}_{2}, \\
S_{\varepsilon_{1}^{\ast},\varepsilon_{2}^{\ast}} &=& 3 ( \vec{\varepsilon}^{\,(\lambda_{1})\ast} \!\cdot\!\hat{r} ) ( \vec{\varepsilon}^{\,(\lambda_{2})\ast} \!\cdot\!\hat{r} ) -  \vec{\varepsilon}^{\,(\lambda_{1})\ast} \!\cdot\! \vec{\varepsilon}^{\,(\lambda_{2})\ast}, \\
S_{\varepsilon_{1}^{\ast},T_{2}} &=& 3 ( \vec{\varepsilon}^{\,(\lambda_{1})\ast} \!\cdot\!\hat{r} ) ( \vec{T}_{2} \!\cdot\!\hat{r} ) -  \vec{\varepsilon}^{\,(\lambda_{1})\ast} \!\cdot\! \vec{T}_{2},
\end{eqnarray} 
where $\hat{r}=\vec{r}/r$ is a unit vector between the two
mesons.
 
The $\rho$ meson exchange potentials are similarly obtained from
the interaction Lagrangians (\ref{eq:vPP})-(\ref{eq:vP*P*}),
\begin{eqnarray}
V^{\rho}_{P_{1}P_{2} \rightarrow P_{1}P_{2}} &\!=\!&
 \left( \frac{\beta g_V}{2m_{\rho}} \right)^{2} \frac{1}{3}  C(r;m_{\rho})
 \vec{\tau}_{1}  \!\cdot\! \vec{\tau}_{2}, \label{eq:rhopot_BBBB}  \\
V^{\rho}_{P_{1}P_{2}^{\ast} \rightarrow 
P_{1}P_{2}^{\ast}} &\!=\!&
 \left( \frac{\beta g_V}{2m_{\rho}} \right)^{2} \frac{1}{3}  C(r;m_{\rho})  \vec{\tau}_{1}  \!\cdot\! \vec{\tau}_{2}, \label{eq:rhopot_BB*BB*}  \\
 V^{\rho}_{P_{1}P_{2}^{\ast} \rightarrow
  P_{1}^{\ast}P_{2}} &\!=\!& 
 \left( 2\lambda g_V \right)^{2} \frac{1}{3} \left[
  2\vec{\varepsilon}_{1}^{\,\ast} \!\cdot\! \vec{\varepsilon}_{2} \,
  C(r;m_{\rho}) \!-\! S_{\varepsilon_{1}^{\ast},\varepsilon_{2}} \, T(r;m_{\rho})
					     \right]  \vec{\tau}_{1}
 \!\cdot\! \vec{\tau}_{2}, \label{eq:rhopot_BB*B*B}  \\
V^{\rho}_{P_{1}^{\ast}P_{2}^{\ast} \rightarrow P_{1}^{\ast}P_{2}^{\ast}} &\!=\!&
 \left(  2\lambda g_V \right)^{2} \frac{1}{3} \left[ 2\vec{T}_{1}
  \!\cdot\! \vec{T}_{2} \, C(r;m_{\rho}) \!-\! S_{T_{1},T_{2}} \,
  T(r;m_{\rho}) \right]  \vec{\tau}_{1}  \!\cdot\! \vec{\tau}_{2} \nonumber \\ 
&& + \left( \frac{\beta g_V}{2m_{\rho}} \right)^{2} \frac{1}{3}  C(r;m_{\rho})
 \vec{\tau}_{1}  \!\cdot\! \vec{\tau}_{2},  \label{eq:rhopot_B*B*B*B*} \\
 V^{\rho}_{P_{1}P_{2} \rightarrow P_{1}^{\ast}P_{2}^{\ast}} &\!=\!&
 \left(  2\lambda g_V  \right)^{2} \frac{1}{3} \left[2
  \vec{\varepsilon}_{1}^{\,\ast} \!\cdot\! \vec{\varepsilon}_{2}^{\,
  \ast} \, C(r;m_{\rho}) \!-\!
  S_{\varepsilon_{1}^{\ast},\varepsilon_{2}^{\ast}} \, T(r;m_{\rho})
					       \right] \vec{\tau}_{1}
 \!\cdot\! \vec{\tau}_{2}, \label{eq:rhopot_BBB*B*} \\
V^{\rho}_{P_{1}P_{2}^{\ast} \rightarrow P_{1}^{\ast}P_{2}^{\ast}} &\!=\!&
 -\left(  2\lambda g_V \right)^{2} \frac{1}{3} \left[ 2\vec{\varepsilon}_{1}^{\,\ast} \!\cdot\! \vec{T}_{2} \, C(r;m_{\rho}) \!-\! S_{\varepsilon_{1}^{\ast},T_{2}} \, T(r;m_{\rho}) \right]  \vec{\tau}_{1}  \!\cdot\! \vec{\tau}_{2}. \label{eq:rhopot_BB*B*B*}
\end{eqnarray}
The $\omega$ meson exchange potentials are obtained by replacing 
the mass of $\rho$ meson with the one of $\omega$ meson and by removing the 
isospin factor $\vec{\tau}_{1}  \!\cdot\! \vec{\tau}_{2}$.
The OPEP's of $\mathrm{P}^{(\ast)}\mathrm{P}^{(\ast)}$ differ from the
ones of $\mathrm{P}^{(\ast)}\bar{\mathrm{P}}^{(\ast)}$ in that 
the overall signs are changed due to $G$-parity. 
The situation is the same with $\omega$ meson exchange potentials,
while $\rho$ meson exchange potentials of $\mathrm{P}^{(\ast)}\bar{\mathrm{P}}^{(\ast)}$ 
are not changed because the $G$-parity is even~\cite{Ohkoda:2011vj}.

In the above equations, $C(r;m_{h})$ and $T(r;m_{h})$  are defined as
\begin{eqnarray}
\hspace{-3em}&&C(r;m_{h}) \!=\! \int \frac{\mbox{d}^{3}\vec{q}}{(2\pi)^3} \frac{m_{h}^{2}}{\vec{q}^{\,\,2}+m_{h}^{2}} 
 e^{i\vec{q} \cdot \vec{r}} \, 
F(\vec{q};m_{h}), \\
\hspace{-3em}&&T(r;m_{h}) S_{12}(\hat{r}) \!=\! \int \frac{\mbox{d}^{3}\vec{q}}{(2\pi)^3} \frac{- \vec{q}^{\,\,2}}{\vec{q}^{\,\,2}+m_{h}^{2}} 
S_{12}(\hat{q})e^{i\vec{q} \cdot \vec{r}} F(\vec{q};m_{h}),
\end{eqnarray}
with $S_{12}(\hat{x}) \!=\! 3 (\vec{\sigma}_{1} \!\cdot\! \hat{x})
(\vec{\sigma}_{2} \!\cdot\! \hat{x}) - \vec{\sigma}_{1} \!\cdot\!
\vec{\sigma}_{2}$.
We introduce the monopole type form factor at each vertex to take into account of 
the size effect of $\mathrm{P}^{(*)}$ mesons.
Then the function reflected form factors is defined as
\begin{eqnarray}
F(\vec{q};m_{h})\!=\! \left(\frac{\Lambda_{\mathrm{P}}^{2} \!-\!
m_{h}^{2}}{\Lambda_{\mathrm{P}}^{2} \!+\! \vec{q}^{\,\,2}} \right)^2 \, ,
\end{eqnarray}
where $m_h$ and $\vec{q}$ are the mass and three-momentum of 
the exchanged meson $h$ ($= \pi, \rho, \omega$) and $\Lambda_{\mathrm{P}}$ is the
cut-off parameter.
The cut-off parameter $\Lambda_{\mathrm{P}}$ are determined from the size of
$\mathrm{P}$ estimated from the constituent quark model as discussed in Refs.~\cite{Yasui:2009bz,Yamaguchi:2011xb,Ohkoda:2011vj,Yamaguchi:2011qw}.
The cut-off parameters are $\Lambda_\mathrm{D} = 1121$ MeV and $\Lambda_\mathrm{B} = 1070$
MeV when the $\pi$ exchange potential is employed, 
while $\Lambda_\mathrm{D} = 1142$ MeV and $\Lambda_\mathrm{B} = 1091$ MeV when
the $\pi \rho \, \omega$ is employed.

Up to now we have given the meson-exchange potentials between two $\mathrm{P}^{(\ast)}$ mesons.
We should note that the potentials contain spin operators and tensor operators, hence that the potentials for each quantum number are different.
In the next section, we classify all the $\mathrm{P}^{(\ast)}\mathrm{P}^{(\ast)}$ states up to $J=2$ and give the corresponding potentials in matrix forms. 

\section{Bound and resonant states}

Let us classify all the possible quantum numbers of the $\mathrm{P}^{(\ast)}\mathrm{P}^{(\ast)}$ systems with isospin $I$, total angular momentum $J$ ($J \le 2$), and parity $P$.
We need also principal quantum number $n=0$, $1$, $\dots$, if there
exist several  bound states for a given $I(J^{P})$.
We show the quantum numbers $I({J^P})$ and the channels in the wave functions in Table~\ref{tbl:classification}. 
It is noted that the wave functions must be symmetric under the exchange of the two $\mathrm{P}^{(\ast)}$ mesons.
We use the notation $^{2S+1}L_J$ to indicate the states with the 
internal spins $S$ and angular momentum $L$.
For example, the $I(J^{P})=0(1^+)$ state is a superposition of four channels; $\frac{1}{\sqrt{2}} \left( \mathrm{P}\mathrm{P}^{\ast}-\mathrm{P}^{\ast}\mathrm{P} \right) (^{3}S_{1})$, $\frac{1}{\sqrt{2}} \left( \mathrm{P}\mathrm{P}^{\ast}-\mathrm{P}^{\ast}\mathrm{P} \right) (^{3}D_{1})$, $\mathrm{P}^{\ast}\mathrm{P}^{\ast}(^{3}S_{1})$ and $\mathrm{P}^{\ast}\mathrm{P}^{\ast}(^{3}D_{1})$.
All the possible channels should be mixed for a given the quantum number.
In the previous studies, the channel mixings were not fully considered \cite{Tornqvist:1993ng,Ding:2009vj,Molina:2010tx}.
Here we pay an attention to that the approximate mass degeneracy of $\mathrm{P}$ and $\mathrm{P}^{\ast}$ plays a crucial role to mix the channels.
Otherwise
 the attraction from the mixing effect becomes suppressed.
We note that the tensor force is also important to mix the channels with different angular momenta, $L$ and $L\pm2$.
As a result, we obtain  the Hamiltonian in a matrix
form with the basis of those coupled channels.
The explicit forms of the Hamiltonian for each $I(J^{P})$ are summarized in Appendix~A. 

Now we are ready to solve the coupled channel Schr\"odinger equations for each quantum number.
The renormalized Numerov method~\cite{johnson} is adopted to numerically
solve the coupled second-order differential equations. 
The resonant states are identified by the behavior of  the phase shift $\delta$ as a
function of the scattering energy $E$. The resonance position $E_r$ 
is defined by an inflection point of the phase shift $\delta(E)$ and the
resonance width by $\Gamma_r = 2/(d\delta /dE)_{E=E_r}$  following Refs.~\cite{Arai:1999pg,Yamaguchi:2011xb,Ohkoda:2011vj}.
To check the consistency of our numerical calculations,
we also adopt the complex scaling method (CSM), in which the resonant state is defined as a pole in the complex energy plane \cite{Arai:1999pg,CSM}.
We obtain an agreement in the results of the renormalized Nemerov method and the CSM.

\begin{figure}[htbp]
\includegraphics[width=13cm]{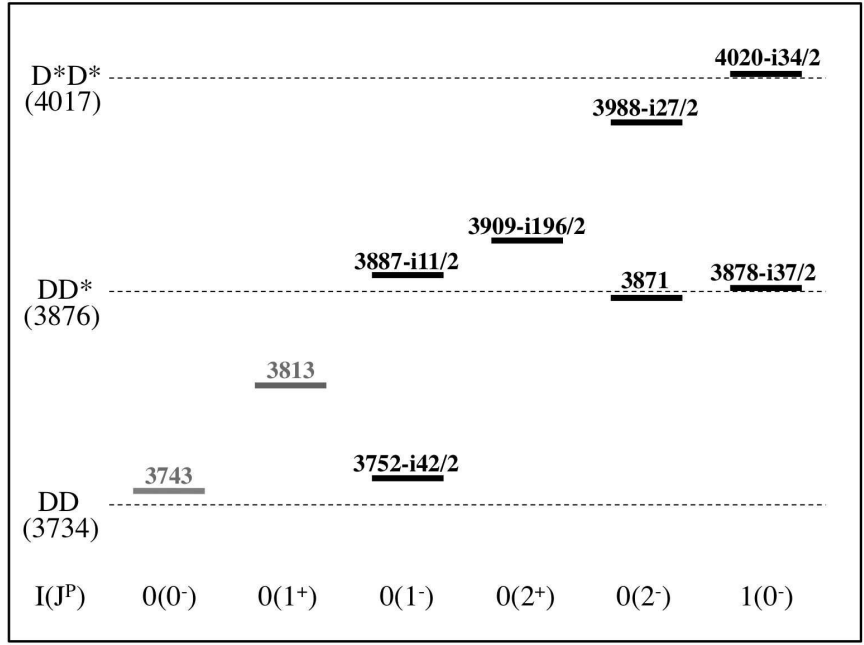}
\caption{Masses of $\mathrm{D}^{(*)}\mathrm{D}^{(*)}$ bound and
 resonant states for various $I(J^{P})$. Solid lines are for our
 predictions with numerical values as denoted above the lines,  and the values in
 parentheses below the lines denote the decay width $\Gamma$ of the resonances when the $\pi \rho\, \omega$
 potential is employed. Mass values are given in units of MeV.
Compact objects which can not be regarded as molecular states are shown
with grey lines.}
\label{fig:resultDD}
\end{figure}

We summarize our numerical results for $\mathrm{D}^{(*)}\mathrm{D}^{(*)}$ bound/resonant
states in Table~\ref{tbl:result_table_DD} and Fig~\ref{fig:resultDD}.
In $\mathrm{D}^{(*)}\mathrm{D}^{(*)}$ states, we find several bound and/or resonant states in $I=0$,
while there is no bound state in $I=1$.
In general, the attractive force of pion exchange in $I=1$ is three times
weaker than in $I=0$ due to the isospin factor.
As a numerical result, $\mathrm{D}^{(*)}\mathrm{D}^{(*)}$ bound states in $I=1$
are not obtained but only resonant states are.

Let us look at our results one by one for each quantum number in detail.
In the following text, most of the numerical values are those for the
case of  the $\pi\rho\,\omega$ potential, because the results from the
$\pi\rho\,\omega$ potential are generally not so different from those
from the $\pi$ potential, except for the $0(2^{-})$ state.
The energies are measured from the threshold,
which is defined to be the lowest mass among the channels for a given
quantum number as tabulated in Table~\ref{tbl:classification}.
For example, we adopt the $\mathrm{D}\mathrm{D}^{\ast}$ mass as threshold for
$I(J^P) =0(0^-)$, while, the  $\mathrm{D}\mathrm{D}$  mass for $I(J^P)=0(1^-)$.

\begin{itemize}
 \item[$0(0^-)$] This state has only one channel of 
	      $\mathrm{D}\mathrm{D}^{\ast}$ 
	      (see Table~.\ref{tbl:classification}) and the pion exchange
	      potential is attractive as shown in Eq.~(\ref{eq:pi_pot_00-}).
	      As a result, the very deep bound state of
	      $\mathrm{D}\mathrm{D}^{\ast}$ is generated with 
	      binding energy 132.1 MeV measured from the $\mathrm{D}\mathrm{D}^{\ast}$ threshold.
 \item[$0(1^+)$] The pion exchange potential is repulsive for diagonal 
	      components as shown in Eq.~(\ref{eq:pi_pot_1+-}). However this state has four components and
	      the mixing of the $S$- and $D$-waves causes the strong tensor
	      attraction from the off-diagonal components of the
	      potential. Consequently, there is a deeply bound state  of
	      mostly $\mathrm{D}\mathrm{D}^{\ast}$ with
	      binding energy 62.3 MeV measured from the $\mathrm{D}\mathrm{D}^{\ast}$ threshold.
 \item[$0(1^-)$] There are twin shape resonances of
	      $\mathrm{D}\mathrm{D}$ with the resonance  
	      energy 17.8 MeV and the decay width 41.6 MeV for the 
	      first resonance, and the resonance energy 152.8 MeV and 
	      the decay width 10.6 MeV for the second. The resonance
	      energies are measured from the $\mathrm{D}\mathrm{D}$
	      threshold. Those resonances are formed by the centrifugal barrier in the $P$-wave.
 \item[$0(2^+)$] This state contains only $D$-wave components of
	      $\mathrm{D}\mathrm{D}^{\ast}$ and $\mathrm{D}^{\ast}\mathrm{D}^{\ast}$.
	      The potential is weakly attractive. 
	      Nevertheless, due to the centrifugal barrier in the
	      $D$-wave, there is a shape resonance of
	      $\mathrm{D}\mathrm{D}^{\ast}$ scattering at the energy 33.7 MeV from the $\mathrm{D}\mathrm{D}^{\ast}$ threshold, but the decay width 196.3 MeV is very wide.
 \item[$0(2^-)$] When the OPEP is employed, there 
	      are twin resonant states with the resonance energy 0.1 MeV
	      and the decay width 0.02 MeV for the first resonance, and the
	      resonance energy 118.0 MeV and the decay width 23.4 MeV
	      for the second from the $\mathrm{D}\mathrm{D}^{\ast}$ threshold.
	      When the effects of $\rho$ and $\omega$ meson exchange are
	      included, 
	      the first resonance becomes  a weakly bound state with
	      the binding energy 4.3 MeV, 
	      because the  $\rho$ meson exchange enhances the central
	      force attraction of the pion exchange.  The $\omega$ meson exchange plays
	      a minor role due to the isospin factor, although this
	      contribution suppresses the attractive central force. The
	      second resonant state with the resonance energy 112.1 MeV
	      and the decay width 26.6 MeV is not affected very much.
 From the analysis of wave function components of the two
	      bound states, we have verified that the lower and higher
	      states are dominated by $\mathrm{D}\mathrm{D}^{\ast}$ and
	      $\mathrm{D}^{\ast}\mathrm{D}^{\ast}$, respectively. 
 \item[$1(0^-)$] This is the only $I=1$ state in $\mathrm{D}^{(\ast)}\mathrm{D}^{(\ast)}$. The interaction in $I=1$ are either repulsive or only
	      weakly attractive as already discussed. Nevertheless, due to the $P$-wave centrifugal barrier, we  find twin shape
	      resonances; the resonance  
	      energy 2.3 MeV and the decay width 37.4 MeV for the 
	      first resonance, and the resonance energy 144.2 MeV and 
	      the decay width 34.4 MeV for the second, from the $\mathrm{D}\mathrm{D}^{\ast}$ threshold.
	    
\end{itemize}

Here several comments are in order.
First, we have obtained several bound and/or resonant state even for $J=2$.
Here the long range force by the OPEP becomes effective for the extended objects with large angular momenta.
It is also interesting to have ``twin states" for several quantum numbers, $0(1^{-})$, $0(2^{-})$ and $1(0^{-})$.
We have to note that the channel couplings by $\mathrm{D}$ and $\mathrm{D}^{\ast}$ are important to produce the obtained energy spectrum.
Indeed, if we cut the channel coupling, we confirm that many of the states disappear.
Thus, we consider that the pattern of the energy spectrum is reflected by the dynamics of the fully coupled channels.

Second, the present formalism of the hadronic molecule picture cannot be applied to compact objects.
When the two $\mathrm{P}^{(\ast)}$ mesons overlap spatially, we need to
consider the internal structure of $\mathrm{P}^{(\ast)}$, which is not
included in the present hadronic picture.
Therefore we shall adopt 1 fm or larger for the size of the bound state to
be interpreted as a molecular state. 
The size of 1 fm is twice of typical radius of $\mathrm{P}^{(\ast)} \sim 0.5$ fm.
For instance, the $I(J^P) = 0(2^-)$ bound state with the binding energy $-4.3$ Mev
is identified with a molecular state because it has the size $1.6$ fm, 
while  the $I(J^P) = 0(0^-)$ bound state with $-132.1$ MeV is not because its size is $0.8$ fm.
We emphasize, however, that this criterion is not definitive but gives only qualitative guide.

\begin{figure}[htbp]
  \includegraphics[width=13cm]{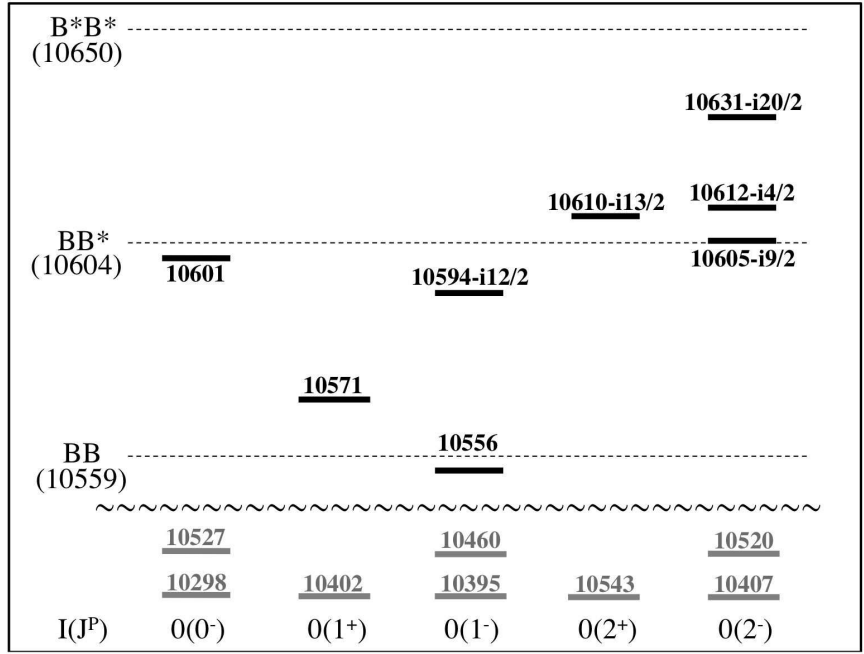}
\caption{ The $\mathrm{B}^{(*)}\mathrm{B}^{(*)}$ bound and
 resonant states around the thresholds with $I(J^{P})$ in $I=0$.
Compact objects which can not be regarded as molecular states are shown
 below the wavy line, but their location do not reflect correct energy scale.
(Same convention as Fig.~\ref{fig:resultDD}.)}
\label{fig:resultBB0}
\end{figure}

\begin{figure}[htbp]
  \includegraphics[width=13cm]{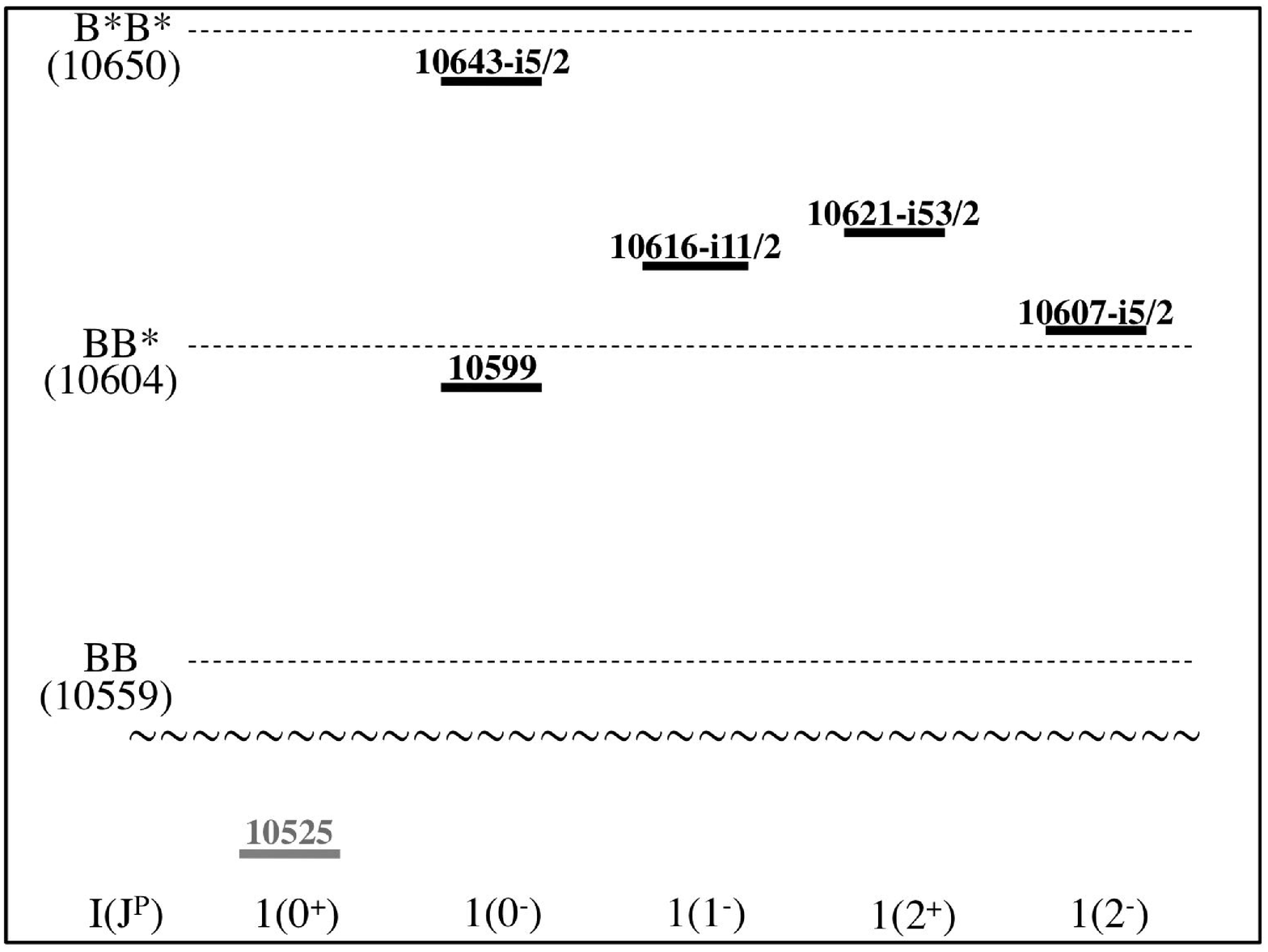}
\caption{ The $\mathrm{B}^{(*)}\mathrm{B}^{(*)}$ bound and
 resonant states around the thresholds with $I(J^{P})$ in $I=1$.
(Same convention as Fig.~\ref{fig:resultDD}.)}
\label{fig:resultBB1}
\end{figure}

Next we discuss the $\mathrm{B}^{(*)}\mathrm{B}^{(*)}$ states.
We use the same coupling constants,
and change only the masses of heavy mesons with small difference of the
cutoff parameters.
The results are summarized in Table~\ref{tbl:result_table_BB} and Figs.~\ref{fig:resultBB0} and
\ref{fig:resultBB1}.
%
At first glance, we find that the $\mathrm{B}^{(*)}\mathrm{B}^{(*)}$ states have many bound and resonant
states in comparison with the $\mathrm{D}^{(*)}\mathrm{D}^{(*)}$ states.
There are two reasons.
First, the kinetic term is suppressed in the Hamiltonian
because the reduced mass becomes larger in the bottom sector.
Second, the effect of channel-couplings becomes more important,
 because a pseudoscalar meson $\mathrm{B}$ and a vector meson $\mathrm{B}^*$
become more degenerate thanks to the heavy quark symmetry; similar discussion has
been done in Refs.~\cite{Yasui:2009bz,Yamaguchi:2011xb,Ohkoda:2011vj}.
We note that, as a consequence of the strong attraction, several new states appear in the $\mathrm{B}^{(*)}\mathrm{B}^{(*)}$ states in isospin triplet; $I(J^{P})=1(0^{+})$, $1(1^{-})$, $1(2^{+})$ and $1(2^{-})$.
The corresponding states are not obtained in the $\mathrm{D}^{(*)}\mathrm{D}^{(*)}$ states.

As noted in the charm sector, the deeply bound states will not be within the scope of the hadronic molecule picture.
In the bottom sector, due to more attraction, more bound states are generated.
For example, we have three states ($n=0$, $1$, $2$) in the $0(0^{-})$ state.
However, the $n=0$, $1$ states are too compact (0.5 fm and 0.9 fm, respectively) to be considered as hadronic molecules.
The $n=2$ state is an extended object (2.5 fm), and hence can be considered as the hadronic molecule.
Similarly, it would not be conclusive yet to consider hadronic molecules for the following states with radii less than 1 fm; the $n=0$ state in $0(1^{+})$, the $n=0$, 1 states in $0(1^{-})$, the $n=0$ state in $0(2^{+})$, the $n=0$, 1 states in $0(2^{-})$ and the $n=0$ state in $1(0^{+})$.

\begin{table}[htdp]
\caption{\small Possible channels of $\mathrm{P}^{(\ast)}\mathrm{P}^{(\ast)}(^
{2S+1}L_{J})$ for a set of quantum numbers $I$ and $J^{P}$ for $J \le 2$.}
\begin{center}
{\renewcommand\arraystretch{1.5}
\begin{tabular}{|c|c|c|}
\hline
$I$ & $J^{P}$ & components \\
\hline
 & $0^{-}$ & $\frac{1}{\sqrt{2}}( \mathrm{P}\mathrm{P}^{\ast} + \mathrm{P}^{\ast}\mathrm{P} )(^{3}P_{0})$ \\
\cline{2-3}
 & $1^{+}$ & $\frac{1}{\sqrt{2}} \left( \mathrm{P}\mathrm{P}^{\ast}-\mathrm{P}^{\ast}\mathrm{P} \right) (^{3}S_{1})$, $\frac{1}{\sqrt{2}} \left( \mathrm{P}\mathrm{P}^{\ast}-\mathrm{P}^{\ast}\mathrm{P} \right) (^{3}D_{1})$, $\mathrm{P}^{\ast}\mathrm{P}^{\ast}(^{3}S_{1})$, $\mathrm{P}^{\ast}\mathrm{P}^{\ast}(^{3}D_{1})$ \\
\cline{2-3}
0 & $1^{-}$ & $\mathrm{P}\mathrm{P}(^{1}P_{1})$, $\frac{1}{\sqrt{2}} \left( \mathrm{P}\mathrm{P}^{\ast}+\mathrm{P}^{\ast}\mathrm{P} \right)(^{3}P_{1})$, $\mathrm{P}^{\ast}\mathrm{P}^{\ast}(^{1}P_{1})$, $\mathrm{P}^{\ast}\mathrm{P}^{\ast}(^{5}P_{1})$, $\mathrm{P}^{\ast}\mathrm{P}^{\ast}(^{5}F_{1})$ \\
\cline{2-3}
 & $2^{+}$ & $\frac{1}{\sqrt{2}} \left( \mathrm{P}\mathrm{P}^{\ast}-\mathrm{P}^{\ast}\mathrm{P} \right)(^{3}D_{2})$, $\mathrm{P}^{\ast}\mathrm{P}^{\ast}(^{3}D_{2})$ \\
\cline{2-3}
 & $2^{-}$ & $\frac{1}{\sqrt{2}} \left( \mathrm{P}\mathrm{P}^{\ast}+\mathrm{P}^{\ast}\mathrm{P} \right)(^{3}P_{2})$, $\frac{1}{\sqrt{2}} \left( \mathrm{P}\mathrm{P}^{\ast}+\mathrm{P}^{\ast}\mathrm{P} \right)(^{3}F_{2})$, $\mathrm{P}^{\ast}\mathrm{P}^{\ast}(^{5}P_{2})$, $\mathrm{P}^{\ast}\mathrm{P}^{\ast}(^{5}F_{2})$ \\
\hline
 & $0^{+}$ & $\mathrm{P}\mathrm{P}(^{1}S_{0})$, $\mathrm{P}^{\ast}\mathrm{P}^{\ast}(^{1}S_{0})$, $\mathrm{P}^{\ast}\mathrm{P}^{\ast}(^{5}D_{0})$ \\
 \cline{2-3}
 & $0^{-}$ & $\frac{1}{\sqrt{2}} \left( \mathrm{P}\mathrm{P}^{\ast}-\mathrm{P}^{\ast}\mathrm{P} \right)(^{3}P_{0})$, $\mathrm{P}^{\ast}\mathrm{P}^{\ast}(^{3}P_{0})$ \\
 \cline{2-3}
1 & $1^{+}$ & $\frac{1}{\sqrt{2}} \left( \mathrm{P}\mathrm{P}^{\ast}+\mathrm{P}^{\ast}\mathrm{P} \right) (^{3}S_{1})$, $\frac{1}{\sqrt{2}} \left( \mathrm{P}\mathrm{P}^{\ast}+\mathrm{P}^{\ast}\mathrm{P} \right)(^{3}D_{1})$, $\mathrm{P}^{\ast}\mathrm{P}^{\ast}(^{5}D_{1})$ \\
\cline{2-3}
 & $1^{-}$ & $\frac{1}{\sqrt{2}} \left( \mathrm{P}\mathrm{P}^{\ast}-\mathrm{P}^{\ast}\mathrm{P} \right)(^{3}P_{1})$, $\mathrm{P}^{\ast}\mathrm{P}^{\ast}(^{3}P_{1})$ \\
 \cline{2-3}
 & $2^{+}$ & $\mathrm{P}\mathrm{P}(^{1}D_{2})$, $\frac{1}{\sqrt{2}} \left( \mathrm{P}\mathrm{P}^{\ast}+\mathrm{P}^{\ast}\mathrm{P} \right)(^{3}D_{2})$, $\mathrm{P}^{\ast}\mathrm{P}^{\ast}(^{1}D_{2})$, $\mathrm{P}^{\ast}\mathrm{P}^{\ast}(^{5}S_{2})$, $\mathrm{P}^{\ast}\mathrm{P}^{\ast}(^{5}D_{2})$, $\mathrm{P}^{\ast}\mathrm{P}^{\ast}(^{5}G_{2})$ \\
\cline{2-3}
 & $2^{-}$ & $\frac{1}{\sqrt{2}} \left( \mathrm{P}\mathrm{P}^{\ast}-\mathrm{P}^{\ast}\mathrm{P} \right)(^{3}P_{2})$, $\frac{1}{\sqrt{2}} \left( \mathrm{P}\mathrm{P}^{\ast}-\mathrm{P}^{\ast}\mathrm{P} \right)(^{3}F_{2})$, $\mathrm{P}^{\ast}\mathrm{P}^{\ast}(^{3}P_{2})$, $\mathrm{P}^{\ast}\mathrm{P}^{\ast}(^{3}F_{2})$ \\ 
\hline
\end{tabular}
}
\end{center}
\label{tbl:classification}
\end{table}%

\begin{table}[htdp]
\caption{\small The energies of $\mathrm{D}^{(\ast)}\mathrm{D}^{(\ast)}$ 
states with $I(J^{P})$ with $J \le 2$. The energies $E$ 
 can be either pure real for bound states or
 complex for resonances. The real parts are measured from the thresholds
 as indicated in the third columns. The imaginary parts are half of the
 decay widths of the resonances, $\Gamma/2$. The values in the
 parentheses for the bound states are matter radii (relative distance of
 the two constituents) in units of fm.}
\begin{center}
{\renewcommand\arraystretch{0.8}
\begin{tabular}{|c|c|c|c|c|}
\hline
$I$ & $J^{P}$ & threshold &  \multicolumn{2}{c|}{$E$ [MeV]}  \\
\cline{4-5}
 & &  & $\pi$-potential & $\pi$, $\rho$, $\omega$-potential   \\
\cline{1-5}
 & $0^{-}$ & $\mathrm{D}\mathrm{D}^{\ast}$ & $-50.2 \;  \; (1.0)$ &
 $-132.1 \; \; (0.8)$ \\
\cline{2-5}
 & $1^{+}$ & $\mathrm{D}\mathrm{D}^{\ast}$ & $-45.7 \;  \; (0.9)$ &
 $-62.3 \;  \; (0.8)$ \\
\cline{2-5}
0 & \multirow{2}*{$1^{-}$} & \multirow{2}*{ $\mathrm{D}\mathrm{D}$ } &
	       $175.4-i\frac{37.4}{2}$ &  $152.8-i\frac{10.6}{2}$ \\
& & &  $19.4-i\frac{63.0}{2}$  & $17.8-i\frac{41.6}{2}$\\
\cline{2-5}
 & $2^{+}$ & $\mathrm{D}\mathrm{D}^{\ast}$ & $34.5-i\frac{183.1}{2}$ 
 & $33.7-i\frac{196.3}{2}$ \\
\cline{2-5}
 & \multirow{2}*{$2^{-}$} & \multirow{2}*{$\mathrm{D}\mathrm{D}^{\ast}$} & 
	     $118.0-i\frac{23.4}{2}$ 
 &  $112.1-i\frac{26.6}{2}$\\
& & & $0.1-i\frac{0.02}{2}$ & $-4.3  \;  \; (1.6)$ \\
\hline
 & $0^{+}$ & $\mathrm{D}\mathrm{D}$ & no & no \\
\cline{2-5}
 & \multirow{2}*{$0^{-}$} & \multirow{2}*{$\mathrm{D}\mathrm{D}^{\ast}$} & 
	     $143.8-i\frac{40.0}{2}$  
 & $144.2-i\frac{34.4}{2}$ \\
& & & $3.7-i\frac{27.7}{2}$ &  $2.3-i\frac{37.4}{2}$ \\
\cline{2-5}
1 & $1^{+}$ & $\mathrm{D}\mathrm{D}^{\ast}$ & no  & no\\
\cline{2-5}
 & $1^{-}$ & $\mathrm{D}\mathrm{D}^{\ast}$ & no  & no\\
\cline{2-5}
 & $2^{+}$ & $\mathrm{D}\mathrm{D}$ & $289.4-i\frac{10.9}{2}$ & no \\
\cline{2-5}
 & $2^{-}$ & $\mathrm{D}\mathrm{D}^{\ast}$ & no  & no \\
\hline
\end{tabular}
}
\end{center}
\label{tbl:result_table_DD}
\end{table}%

\begin{table}[htdp]
\caption{\small The energies of $\mathrm{B}^{(\ast)}\mathrm{B}^{(\ast)}$ 
states with $I(J^{P})$ with $J \le 2$. (Same convention as Table~\ref{tbl:result_table_DD}.)}
\begin{center}
{\renewcommand\arraystretch{0.8}
\begin{tabular}{|c|c|c|c|c|}
\hline
$I$ & $J^{P}$ & threshold &  \multicolumn{2}{c|}{$E-i\Gamma/2$ [MeV]} \\
\cline{4-5}
 & &  & $\pi$-potential & $\pi$, $\rho$, $\omega$-potential   \\
\cline{1-5}
 & \multirow{3}*{$0^{-}$} & \multirow{3}*{$\mathrm{B}\mathrm{B}^{\ast}$} &  &
    $-3.3 \; \; (2.5)$ \\
& & & $-32.0 \; \; (1.2)$ & $-77.5 \; \; (0.9)$ \\
& & & $-178.0 \; \; (0.6)$ & $-305.9 \; \; (0.5)$ \\
\cline{2-5}
 & \multirow{2}*{$1^{+}$} & \multirow{2}*{$\mathrm{B}\mathrm{B}^{\ast}$}
	 &  $-25.7 \; \; (1.2)$ 
 & $-33.6 \; \; (1.1)$ \\
& & & $-179.7 \; \; (0.5)$ &  $-201.5 \; \; (0.5)$ \\
\cline{2-5}
\multirow{2}*{0} & \multirow{4}*{$1^{-}$} & \multirow{4}*{$\mathrm{B}\mathrm{B}$} 
& $52.8-i\frac{12.8}{2}$  & $35.0-i\frac{11.8}{2}$  \\
 & & & $1.9-i\frac{3.3}{2}$ & $-3.1 \; \; (1.6)$ \\
 & & & $-39.1 \; \; (0.8)$ & $-98.9 \; \; (0.6)$ \\
 & & & $-125.5 \; \; (0.6)$ & $-164.4 \; \; (0.5)$ \\
\cline{2-5}
 & \multirow{2}*{$2^{+}$} & \multirow{2}*{$\mathrm{B}\mathrm{B}^{\ast}$} &  $5.5-i\frac{5.9}{2}$  
 &  $5.7-i\frac{13.2}{2}$\\
& & & $-51.2 \; \; (0.8)$ & $-60.6 \; \; (0.8)$ \\
\cline{2-5}
 & \multirow{5}*{$2^{-}$} & \multirow{5}*{$\mathrm{B}\mathrm{B}^{\ast}$} 
 &   &  $26.9-i\frac{20.2}{2}$ \\
 & & &  $26.9-i\frac{20.2}{2}$ &   $7.6-i\frac{4.3}{2}$ \\
& & & $7.6-i\frac{4.3}{2}$ & $0.5-i\frac{8.5}{2}$ \\
& & & $-68.7 \; \; (0.8)$ & $-84.1 \; \; (0.7)$ \\
& & & $-147.5 \; \; (0.6)$ & $-196.5 \; \; (0.6)$ \\ 
\hline
 & $0^{+}$ & $\mathrm{B}\mathrm{B}$ & $-18.1 \; \; (0.9)$, & $-33.9 \;
 \; (0.5)$  \\
\cline{2-5}
 & \multirow{3}*{$0^{-}$} & \multirow{2}*{$\mathrm{B}\mathrm{B}^{\ast}$} &   $46.7-
i\frac{1.9}{2}$ &   \\
& & & $0.7-i\frac{3.0}{2}$ & $38.5-i\frac{4.6}{2}$ \\
& & & $-50.5 \; \; (0.8)$ & $-5.9 \; \; (1.4)$ \\
\cline{2-5}
1 & $1^{+}$ & $\mathrm{B}\mathrm{B}^{\ast}$ & $-38.1 \; \; (0.8)$  & no \\
\cline{2-5}
 & $1^{-}$ & $\mathrm{B}\mathrm{B}^{\ast}$ & no & $11.7-i\frac{11.0}{2}$ \\
\cline{2-5}
 & $2^{+}$ & $\mathrm{B}\mathrm{B}$ & $23.0 - i\frac{4.4}{2}$ & $62.4-i\frac{52.5}{2}$  \\
\cline{2-5}
 & \multirow{2}*{$2^{-}$} & \multirow{2}*{$\mathrm{B}\mathrm{B}^{\ast}$} &  $63.7-i\frac{7.6}{2}$  &  \\
& & & $2.0-i\frac{3.3}{2}$ & $2.3-i\frac{4.7}{2}$ \\
\hline
\end{tabular}
}
\end{center}
\label{tbl:result_table_BB}
\end{table}%

\section{Hadronic molecules and tetraquarks}

In the present paper, we have considered $\mathrm{T}_{\mathrm{QQ}}$ as a hadronic molecule composed by $\mathrm{P}^{(\ast)}\mathrm{P}^{(\ast)}$ in which one pion exchange potential induces a dominant attraction.
On the other hand, as mentioned in the introduction,  $\mathrm{T}_{\mathrm{QQ}}$  can also be considered as a tetraquark $\bar{\mathrm{Q}}\bar{\mathrm{Q}}\mathrm{q}\mathrm{q}$ in which a diquark $\mathrm{q}\mathrm{q}$ provides a strong binding energy.
The different features between the hadronic molecule and tetraquark pictures are seen 
in their sizes.  
For hadronic molecules to be valid, hadron constituents are sufficiently 
far apart such that their identities as hadrons must be maintained.  
They can not be too close to overlap each other.  
Therefore, masses of hadronic molecules should appear 
around their threshold regions.  
In contrast, tetraquarks may be strongly bound and become compact objects 
as genuine quark objects.  
Thus, their natures are differentiated in the masses, small binding energy of order ten MeV or less, or larger one.
Although it is tempting to seek for a framework to cover both scales, such a problem is out of scope of the present paper.
Instead, we compare the results from the two pictures and just clarify the differences between them.

In the hadronic molecule picture by $\mathrm{P}^{(\ast)}\mathrm{P}^{(\ast)}$, as presented in the previous section, we obtain, not only the bound states, but also the resonant states in many $I(J^{P})$ quantum numbers.
Both in charm and bottom sectors, it is remarkable that there are even the twin states in several quantum numbers, $0(1^{-})$,  $0(2^{-})$  and  $1(0^{-})$.

In the tetraquark picture including a diquark model \cite{Lee:2007tn,Lee:2009rt,Vijande:2009kj,Vijande:2011zz}, in contrast, only two bound states in $I(J^{P})=0(1^{+})$ and $1(2^{-})$ have been predicted until now, as discussed for an example in Refs.~\cite{Vijande:2009kj,Carames:2011zz,Vijande:2011zz} as recent works.
For $0(1^{+})$, the predicted binding energy can be around 70 MeV from the $\mathrm{D}\mathrm{D}^{\ast}$ threshold. 
For $1(2^{-})$, the predicted binding energy can be around 27 MeV.
The mass of the $0(1^{+})$ states in the tetraquark picture is accidentally close to our value 62 MeV in the hadronic molecule picture.
However, this comparison should be considered more carefully.
Because the size of the $0(1^{+})$ state is 0.8 fm from Table~\ref{tbl:result_table_DD},
the $\mathrm{D}$ and $\mathrm{D}^{\ast}$ mesons composing the $0(1^{+})$ state would be overlapped spatially if the size of each $\mathrm{D}$ and $\mathrm{D}^{\ast}$ meson is a scale of about 1 fm.
In such a compact object, the quark degrees of freedom may become active to contribute to the dynamics like the tetraquark picture.
However, such an effect is out of the scope of the present hadronic molecule picture.
The $1(2^{-})$ state cannot be found in our present study.

In any cases, the feature in the energy spectrum in the $\mathrm{D}^{(\ast)}\mathrm{D}^{(\ast)}$ molecule picture is that there are many shallow bound states and resonances in $0(1^{-})$, $0(2^{+})$, $0(2^{-})$ and $1(0^{-})$,
which are not found in the tetraquark picture.
Thus, we observe that the energy spectrum of the hadronic molecule picture by $\mathrm{P}^{(\ast)}\mathrm{P}^{(\ast)}$ is qualitatively different from that of the tetraquark picture.

\section{Summary}
We have discussed exotic mesons with double charm and bottom flavor whose quark content $\bar{\mathrm Q}\bar{\mathrm Q}\mathrm{q}\mathrm{q}$ is genuinely exotic.
We have taken the hadronic picture, and considered molecular states of two heavy mesons $\mathrm{P}^{(*)}$'s (a pseudoscalar meson $\mathrm{P}=\mathrm{D}$, $\mathrm{B}$, and a vector meson $\mathrm{P}^{\ast}=\mathrm{D}^{\ast}$, $\mathrm{B}^{\ast}$).
With respecting the heavy quark  and chiral
symmetries, we have constructed the $\pi$ exchange potential and the $\pi\rho\,\omega$ exchange potential between the two heavy
mesons.
To investigate the bound and/or resonant states, we have numerically solved the coupled channel Schr\"odinger equations
for the $\mathrm{P}^{(\ast)}\mathrm{P}^{(\ast)}$ states with $I(J^P)$ for $J\leq2$.

As  results, we have found many bound and/or resonant states in both charm and bottom sectors. 
The $\mathrm{D}^{(*)}\mathrm{D}^{(*)}$ bound and 
resonant states have moderate energies and decay widths around the
thresholds in several channels with quantum numbers; $0(0^{-})$, $0(1^{+})$, $0(1^{-})$, $0(2^{+})$, $0(2^{-})$ and $1(0^{-})$.
The $\mathrm{B}^{(*)}\mathrm{B}^{(*)}$ states have more bound
and resonant states with various quantum numbers.
Several new states appear in the $\mathrm{B}^{(*)}\mathrm{B}^{(*)}$
states in isotriplet states, such as $1(0^{+})$, $1(1^{-})$, $1(2^{+})$ and $1(2^{-})$, which cannot be found in the charm sector.
By contrast to the $\mathrm{D}^{(*)}\mathrm{D}^{(*)}$ states, some $\mathrm{B}^{(*)}\mathrm{B}^{(*)}$ states are very compact 
objects with a large binding energy much below the thresholds.
Perhaps, these states cannot survive as hadronic molecules and more 
consideration of quark dynamics such as tetraquarks is required.

The energy spectrum for quantum numbers $I(J^{P})$ will help us to study the structure of the exotic states with $\bar{\mathrm Q}\bar{\mathrm Q}\mathrm{q}\mathrm{q}$.
In the present hadronic molecule picture, many shallow bound states and resonant states appear around the thresholds in several quantum numbers.
Indeed, they were not found in the tetraquark picture.
It is interesting to note that, around the thresholds for $0(1^{-})$, $0(2^{+})$, $0(2^{-})$ and $1(0^{-})$, the shape of energy spectrum in those states in the charm sector seems to that in he bottom sector.
It may indicate a universal behavior of the energy spectrum around the thresholds.

Experimental studies of those exotic hadrons should be performed in the
coming future. The double charm production in accelerator facilities
will help us to search them \cite{DelFabbro:2004ta}. Recently it has
been discussed that the quark-gluon plasma in the relativistic heavy ion
collisions could produce much abundance of exotic hadrons including the
exotic mesons with double charm \cite{Cho:2010db,Cho:2011ew}. Those
experimental studies will shed a light on the nature of the exotic
mesons with double charm and bottom flavor, and provide important hints to the fundamental questions of the strong interaction in hadron physics. 

\section*{Acknowledgments}
We thank  Prof.~S.~Takeuchi and Prof.~M.~Takizawa for fruitful discussions and comments.
This work is supported in part by Grant-in-Aid for Scientific Research on 
Priority Areas ``Elucidation of New Hadrons with a Variety of Flavors 
(E01: 21105006)" (S.Y. and A.H.) and by ``Grant-in-Aid for Young Scientists (B)
22740174" (K.S.), from 
the ministry of Education, Culture, Sports, Science and Technology of
Japan.

\appendix
\section{Hamiltonian}

The Hamiltonian is a sum of the kinetic term and potential term as,
\begin{eqnarray}
H_{I(J^P)} = K_{I(J^P)} + V^{\pi}_{I(J^P)},
\end{eqnarray}
for the $\pi$ exchange potential only, and 
\begin{eqnarray}
H_{I(J^P)} = K_{I(J^P)} + \sum_{i=\pi, \rho, \omega}V^{i}_{I(J^P)},
\end{eqnarray}
for the $\pi \rho\, \omega$ potential.

The kinetic terms with including the explicit breaking of the heavy
quark symmetry by the mass difference $ \Delta
m_{\mathrm{P}\mathrm{P}^{\ast}}=m_{\mathrm{P}^{\ast}}-m_{\mathrm{P}}$ are
\begin{align}
 K_{0(0-)} &=
\mathrm{diag} \left(
 -\frac{1}{2\tilde{m}_{\mathrm{P}\mathrm{P}^{\ast}}} \triangle_{1}
  \right), \\
K_{0(1^+)} &=
\mathrm{diag} \left(
 -\frac{1}{2\tilde{m}_{\mathrm{P}\mathrm{P}^{\ast}}} \triangle_{0},
 -\frac{1}{2\tilde{m}_{\mathrm{P}\mathrm{P}^{\ast}}} \triangle_{2},
 -\frac{1}{2\tilde{m}_{\mathrm{P}^{\ast}\mathrm{P}^{\ast}}} \triangle_{0} + \Delta m_{\mathrm{P}\mathrm{P}^{\ast}},
 -\frac{1}{2\tilde{m}_{\mathrm{P}^{\ast}\mathrm{P}^{\ast}}} \triangle_{2} + \Delta m_{\mathrm{P}\mathrm{P}^{\ast}}
 \right), \\
K_{0(1^-)} &=
\mathrm{diag} \left(
 -\frac{1}{2\tilde{m}_{\mathrm{P}\mathrm{P}}} \triangle_{0},
 -\frac{1}{2\tilde{m}_{\mathrm{P}\mathrm{P}^{\ast}}} \triangle_{1} + \Delta m_{\mathrm{P}\mathrm{P}^{\ast}}, \right. \nonumber \\ 
& \left. -\frac{1}{2\tilde{m}_{\mathrm{P}^{\ast}\mathrm{P}^{\ast}}} \triangle_{1} + 2\Delta m_{\mathrm{P}\mathrm{P}^{\ast}},
 -\frac{1}{2\tilde{m}_{\mathrm{P}^{\ast}\mathrm{P}^{\ast}}} \triangle_{1} + 2\Delta m_{\mathrm{P}\mathrm{P}^{\ast}},
 -\frac{1}{2\tilde{m}_{\mathrm{P}^{\ast}\mathrm{P}^{\ast}}} \triangle_{3} + 2\Delta m_{\mathrm{P}\mathrm{P}^{\ast}}
 \right), \\
K_{0(2^+)} &=
\mathrm{diag} \left(
 -\frac{1}{2\tilde{m}_{\mathrm{P}\mathrm{P}^{\ast}}} \triangle_{2},
 -\frac{1}{2\tilde{m}_{\mathrm{P}^{\ast}\mathrm{P}^{\ast}}} \triangle_{2} + \Delta m_{\mathrm{P}\mathrm{P}^{\ast}}
 \right), \\
K_{0(2^-)} &=
\mathrm{diag} \left(
 -\frac{1}{2\tilde{m}_{\mathrm{P}\mathrm{P}^{\ast}}} \triangle_{1},
 -\frac{1}{2\tilde{m}_{\mathrm{P}\mathrm{P}^{\ast}}} \triangle_{3},
 -\frac{1}{2\tilde{m}_{\mathrm{P}^{\ast}\mathrm{P}^{\ast}}} \triangle_{1} + \Delta m_{\mathrm{P}\mathrm{P}^{\ast}},
 -\frac{1}{2\tilde{m}_{\mathrm{P}^{\ast}\mathrm{P}^{\ast}}} \triangle_{3} + \Delta m_{\mathrm{P}\mathrm{P}^{\ast}}
 \right), \\
K_{1(0^+)} &=
\mathrm{diag} \left(
 -\frac{1}{2\tilde{m}_{\mathrm{P}\mathrm{P}}} \triangle_{0},
 -\frac{1}{2 \tilde{m}_{\mathrm{P}\mathrm{P}^{\ast}}} \triangle_{0} + 2 \Delta m_{\mathrm{P}\mathrm{P}^{\ast}},
 -\frac{1}{2\tilde{m}_{\mathrm{P}\mathrm{P}^{\ast}}} \triangle_{2} + 2 \Delta m_{\mathrm{P}\mathrm{P}^{\ast}}
  \right), \\
K_{1(0^-)} &=
\mathrm{diag} \left(
 -\frac{1}{2\tilde{m}_{\mathrm{P}\mathrm{P}^{\ast}}} \triangle_{1},
 -\frac{1}{2\tilde{m}_{\mathrm{P}^{\ast}\mathrm{P}^{\ast}}} \triangle_{1} + \Delta m_{\mathrm{P}\mathrm{P}^{\ast}}
  \right), \\
K_{1(1^+)} &=
\mathrm{diag} \left(
 -\frac{1}{2\tilde{m}_{\mathrm{P}\mathrm{P}^{\ast}}} \triangle_{0},
 -\frac{1}{2\tilde{m}_{\mathrm{P}\mathrm{P}^{\ast}}} \triangle_{2},
 -\frac{1}{2\tilde{m}_{\mathrm{P}^{\ast}\mathrm{P}^{\ast}}} \triangle_{2} + \Delta m_{\mathrm{P}\mathrm{P}^{\ast}}
 \right), \\
K_{1(1^-)} &=
\mathrm{diag} \left(
 -\frac{1}{2\tilde{m}_{\mathrm{P}\mathrm{P}^{\ast}}} \triangle_{1},
 -\frac{1}{2\tilde{m}_{\mathrm{P}^{\ast}\mathrm{P}^{\ast}}} \triangle_{1} + \Delta m_{\mathrm{P}\mathrm{P}^{\ast}}
 \right), \\
K_{1(2^+)} &=
\mathrm{diag} \left(
 -\frac{1}{2\tilde{m}_{\mathrm{P}\mathrm{P}}} \triangle_{2},
 -\frac{1}{2\tilde{m}_{\mathrm{P}\mathrm{P}^{\ast}}} \triangle_{2} + \Delta m_{\mathrm{P}\mathrm{P}^{\ast}},
 -\frac{1}{2\tilde{m}_{\mathrm{P}^{\ast}\mathrm{P}^{\ast}}} \triangle_{2} + 2 \Delta m_{\mathrm{P}\mathrm{P}^{\ast}}, \right. \nonumber \\
& \left. -\frac{1}{2\tilde{m}_{\mathrm{P}^{\ast}\mathrm{P}^{\ast}}} \triangle_{0} + 2 \Delta m_{\mathrm{P}\mathrm{P}^{\ast}},
 -\frac{1}{2\tilde{m}_{\mathrm{P}^{\ast}\mathrm{P}^{\ast}}} \triangle_{2} + 2 \Delta m_{\mathrm{P}\mathrm{P}^{\ast}},
 -\frac{1}{2\tilde{m}_{\mathrm{P}^{\ast}\mathrm{P}^{\ast}}} \triangle_{4} + 2 \Delta m_{\mathrm{P}\mathrm{P}^{\ast}}
 \right), \\
K_{1(2^-)} &=
\mathrm{diag} \left(
 -\frac{1}{2\tilde{m}_{\mathrm{P}\mathrm{P}^{\ast}}} \triangle_{1},
 -\frac{1}{2\tilde{m}_{\mathrm{P}\mathrm{P}^{\ast}}} \triangle_{3},
 -\frac{1}{2\tilde{m}_{\mathrm{P}^{\ast}\mathrm{P}^{\ast}}} \triangle_{1} + \Delta m_{\mathrm{P}\mathrm{P}^{\ast}},
 -\frac{1}{2\tilde{m}_{\mathrm{P}^{\ast}\mathrm{P}^{\ast}}} \triangle_{3} + \Delta m_{\mathrm{P}\mathrm{P}^{\ast}}
 \right), 
\end{align}
where $\triangle_{l} = \frac{\partial^{2}}{\partial r^{2}}+\frac{2}{r}
\frac{\partial}{\partial r} - \frac{l(l+1)}{r^{2}}$ with integer $l \ge
0$, $1/\tilde{m}_{\mathrm{P}\mathrm{P}} =
1/m_{\mathrm{P}}+1/m_{\mathrm{P}}$,
$1/\tilde{m}_{\mathrm{P}\mathrm{P}^{\ast}} =
1/m_{\mathrm{P}}+1/m_{\mathrm{P}^{\ast}}$,
$1/\tilde{m}_{\mathrm{P}^{\ast}\mathrm{P}^{\ast}} =
1/m_{\mathrm{P}^{\ast}}+1/m_{\mathrm{P}^{\ast}}$.
 
\newpage
The $\pi$ exchange potentials for each $I(J^P)$ states are
{\scriptsize
\begin{eqnarray}
V_{0(0^-)}^{\pi} &=&
\left(
V^{\pi}_{\mathrm C}+2V^{\pi}_{\mathrm T}
\right) \, , \label{eq:pi_pot_00-} \\
V_{0(1+)}^{\pi} &=&
\left(
\begin{array}{cccc}
  -V^{\pi}_{\mathrm C} & \sqrt{2} V^{\pi}_{\mathrm T} & 2V^{\pi}_{\mathrm C} & \sqrt{2} V^{\pi}_{\mathrm T}   \\
  \sqrt{2} V^{\pi}_{\mathrm T} & -V^{\pi}_{\mathrm C}-V^{\pi}_{\mathrm T}  & \sqrt{2} V^{\pi}_{\mathrm T} & 2V^{\pi}_{\mathrm C} - V^{\pi}_{\mathrm T} \\
 2V^{\pi}_{\mathrm C} & \sqrt{2} V^{\pi}_{\mathrm T}  & -V^{\pi}_{\mathrm C} & \sqrt{2} V^{\pi}_{\mathrm T} \\
 \sqrt{2} V^{\pi}_{\mathrm T} & 2V^{\pi}_{\mathrm C} - V^{\pi}_{\mathrm T} & \sqrt{2} V^{\pi}_{\mathrm T} & -V^{\pi}_{\mathrm C} - V^{\pi}_{\mathrm T}
\end{array}
\right) \, ,
\label{eq:pi_pot_1+-} \\
V_{0(1^-)}^{\pi} &=&
\left(
\begin{array}{ccccc}
  0 & 0 & -\sqrt{3} V^{\pi}_{\mathrm C} & -2\sqrt{\frac{3}{5}} V^{\pi}_{\mathrm T} & 3\sqrt{\frac{2}{5}} V^{\pi}_{\mathrm T}   \\
  0 & V^{\pi}_{\mathrm C}-V^{\pi}_{\mathrm T} & 0  & -3\sqrt{\frac{3}{5}} V^{\pi}_{\mathrm T} & -3\sqrt{\frac{2}{5}} V^{\pi}_{\mathrm T} \\
 -\sqrt{3} V^{\pi}_{\mathrm C} & 0 & -2 V^{\pi}_{\mathrm C}  & \frac{2}{\sqrt{5}}V^{\pi}_{\mathrm T} & -\sqrt{\frac{6}{5}} V^{\pi}_{\mathrm T} \\
 -2\sqrt{\frac{3}{5}} V^{\pi}_{\mathrm T} & -3\sqrt{\frac{3}{5}} V^{\pi}_{\mathrm T} & \frac{2}{\sqrt{5}}V^{\pi}_{\mathrm T} & V^{\pi}_{\mathrm C} - \frac{7}{5} V^{\pi}_{\mathrm T} & \frac{\sqrt{6}}{5} V^{\pi}_{\mathrm T} \\
  3\sqrt{\frac{2}{5}} V^{\pi}_{\mathrm T} & -3\sqrt{\frac{2}{5}} V^{\pi}_{\mathrm T} & -\sqrt{\frac{6}{5}} V^{\pi}_{\mathrm T} & \frac{\sqrt{6}}{5} V^{\pi}_{\mathrm T} & V^{\pi}_{\mathrm C} - \frac{8}{5} V^{\pi}_{\mathrm T}
\end{array}
\right) \, ,
\label{eq:pi_pot_1--} \\
V_{0(2^+)}^{\pi} &=&
\left(
\begin{array}{cc}
  -V^{\pi}_{\mathrm C}+V^{\pi}_{\mathrm T} & 2V^{\pi}_{\mathrm C}+V^{\pi}_{\mathrm T}    \\
  2V^{\pi}_{\mathrm C}+V^{\pi}_{\mathrm T} & -V^{\pi}_{\mathrm C}+V^{\pi}_{\mathrm T}
\end{array}
\right) \, ,
\label{eq:pi_pot_2+-} \\
V_{0(2^-)}^{\pi} &=&
\left(
\begin{array}{cccc}
 V^{\pi}_{\mathrm C} + \frac{1}{5} V^{\pi}_{\mathrm T} & -\frac{3\sqrt{6}}{5} V^{\pi}_{\mathrm T} & \frac{3\sqrt{3}}{5} V^{\pi}_{\mathrm T} & -\frac{6\sqrt{3}}{5} V^{\pi}_{\mathrm T}   \\
- \frac{3\sqrt{6}}{5} V^{\pi}_{\mathrm T} & V^{\pi}_{\mathrm C}+\frac{4}{5}V^{\pi}_{\mathrm T}  & \frac{3\sqrt{2}}{5} V^{\pi}_{\mathrm T} & -\frac{6\sqrt{2}}{5} V^{\pi}_{\mathrm T} \\
 \frac{3\sqrt{3}}{5} V^{\pi}_{\mathrm T} & \frac{3\sqrt{2}}{5} V^{\pi}_{\mathrm T}  & V^{\pi}_{\mathrm C} + \frac{7}{5} V^{\pi}_{\mathrm T} & \frac{6}{5} V^{\pi}_{\mathrm T} \\
- \frac{6\sqrt{3}}{5} V^{\pi}_{\mathrm T} & -\frac{6\sqrt{2}}{5} V^{\pi}_{\mathrm T} & \frac{6}{5} V^{\pi}_{\mathrm T} & V^{\pi}_{\mathrm C} - \frac{2}{5} V^{\pi}_{\mathrm T}
\end{array}
\right) \, .
\label{eq:pi_pot_2--} \\
V_{1(0^+)}^{\pi} &=&
\left(
 \begin{array}{ccc}
  0 & -\sqrt{3} V^{\pi}_{\mathrm C} & \sqrt{6} V^{\pi}_{\mathrm T}   \\
  -\sqrt{3} V^{\pi}_{\mathrm C} & -2V^{\pi}_{\mathrm C}  & -\sqrt{2} V^{\pi}_{\mathrm T}  \\
 \sqrt{6} V^{\pi}_{\mathrm T} & -\sqrt{2} V^{\pi}_{\mathrm T}  &  V^{\pi}_{\mathrm C}-2V^{\pi}_{\mathrm T}
\end{array}
\right) \, , 
\label{eq:pi_pot_0++} \\
V_{1(0^-)}^{\pi} &=&
\left(
\begin{array}{cc}
  -V^{\pi}_{\mathrm C}-2V^{\pi}_{\mathrm T} & 2V^{\pi}_{\mathrm C}-2V^{\pi}_{\mathrm T}    \\
  2V^{\pi}_{\mathrm C}-2V^{\pi}_{\mathrm T} & -V^{\pi}_{\mathrm C}-2V^{\pi}_{\mathrm T}
\end{array}
\right) \, ,
\label{eq:pi_pot_0-+} \\
V_{1(1^+)}^{\pi} &=&
\left(
\begin{array}{ccc}
  V^{\pi}_{\mathrm C} & -\sqrt{2} V^{\pi}_{\mathrm T} & -\sqrt{6} V^{\pi}_{\mathrm T}   \\
  -\sqrt{2} V^{\pi}_{\mathrm T} & V^{\pi}_{\mathrm C}+V^{\pi}_{\mathrm T} & -\sqrt{3} V^{\pi}_{\mathrm T}  \\
 -\sqrt{6} V^{\pi}_{\mathrm T} & -\sqrt{3} V^{\pi}_{\mathrm T} &  V^{\pi}_{\mathrm C}-V^{\pi}_{\mathrm T}
\end{array}
\right) \, ,
\label{eq:pi_pot_1++} \\
V_{1(1^-)}^{\pi} &=&
\left(
\begin{array}{cc}
  -V^{\pi}_{\mathrm C}+V^{\pi}_{\mathrm T} & -2V^{\pi}_{\mathrm C}-V^{\pi}_{\mathrm T}    \\
  -2V^{\pi}_{\mathrm C}-V^{\pi}_{\mathrm T} & -V^{\pi}_{\mathrm C}+V^{\pi}_{\mathrm T}
\end{array}
\right) \, ,
\label{eq:pi_pot_1-+} \\
V_{1(2^+)}^{\pi} &=&
\left(
\begin{array}{cccccc}
  0 & 0 & -\sqrt{3} V^{\pi}_{\mathrm C} & \sqrt{\frac{6}{5}} V^{\pi}_{\mathrm T} & -2\sqrt{\frac{3}{7}} V^{\pi}_{\mathrm T} & 6\sqrt{\frac{3}{35}} V^{\pi}_{\mathrm T} \\
  0 & V^{\pi}_{\mathrm C}-V^{\pi}_{\mathrm T} & 0 & 3\sqrt{\frac{2}{5}} V^{\pi}_{\mathrm T} & -\frac{3}{\sqrt{7}} V^{\pi}_{\mathrm T} & -\frac{12}{\sqrt{35}} V^{\pi}_{\mathrm T} \\
  -\sqrt{3} V^{\pi}_{\mathrm C} & 0 & -2 V^{\pi}_{\mathrm C} & -\sqrt{\frac{2}{5}} V^{\pi}_{\mathrm T} & \frac{2}{\sqrt{7}} V^{\pi}_{\mathrm T} & -\frac{6}{\sqrt{35}} V^{\pi}_{\mathrm T} \\
  \sqrt{\frac{6}{5}} V^{\pi}_{\mathrm T} & 3\sqrt{\frac{3}{5}} V^{\pi}_{\mathrm T} & -\sqrt{\frac{2}{5}} V^{\pi}_{\mathrm T} & V^{\pi}_{\mathrm C} & \sqrt{\frac{14}{5}} V^{\pi}_{\mathrm T} & 0 \\
  -2\sqrt{\frac{3}{7}} V^{\pi}_{\mathrm T} & -\frac{3}{\sqrt{7}} V^{\pi}_{\mathrm T} & \frac{2}{\sqrt{7}} V^{\pi}_{\mathrm T} & \sqrt{\frac{14}{5}} V^{\pi}_{\mathrm T} & V^{\pi}_{\mathrm C} + \frac{3}{7} V^{\pi}_{\mathrm T} & \frac{12}{7\sqrt{5}} V^{\pi}_{\mathrm T} \\
  6\sqrt{\frac{3}{35}} V^{\pi}_{\mathrm T} & -\frac{12}{\sqrt{35}} V^{\pi}_{\mathrm T} & -\frac{6}{\sqrt{35}} V^{\pi}_{\mathrm T} & 0 & \frac{12}{7\sqrt{5}} V^{\pi}_{\mathrm T} & V^{\pi}_{\mathrm C} - \frac{10}{7} V^{\pi}_{\mathrm T}
\end{array}
\right) \, ,
\label{eq:pi_pot_2++} \\
V_{1(2^-)}^{\pi} &=&
\left(
\begin{array}{cccc}
  -V^{\pi}_{\mathrm C} - \frac{1}{5} V^{\pi}_{\mathrm T} & \frac{3\sqrt{6}}{5} V^{\pi}_{\mathrm T} & 2V^{\pi}_{\mathrm C} - \frac{1}{5}V^{\pi}_{\mathrm T} & \frac{3\sqrt{6}}{5} V^{\pi}_{\mathrm T}   \\
  \frac{3\sqrt{6}}{5} V^{\pi}_{\mathrm T} & -V^{\pi}_{\mathrm C}-\frac{4}{5}V^{\pi}_{\mathrm T}  & \frac{3\sqrt{6}}{5} V^{\pi}_{\mathrm T} & 2V^{\pi}_{\mathrm C} - \frac{4}{5} V^{\pi}_{\mathrm T} \\
 2V^{\pi}_{\mathrm C} - \frac{1}{5} V^{\pi}_{\mathrm T} & \frac{3\sqrt{6}}{5} V^{\pi}_{\mathrm T}  & -V^{\pi}_{\mathrm C} - \frac{1}{5} V^{\pi}_{\mathrm T} & \frac{3\sqrt{6}}{5} V^{\pi}_{\mathrm T} \\
 \frac{3\sqrt{6}}{5} V^{\pi}_{\mathrm T} & 2V^{\pi}_{\mathrm C} - \frac{4}{5} V^{\pi}_{\mathrm T} & \frac{3\sqrt{6}}{5} V^{\pi}_{\mathrm T} & -V^{\pi}_{\mathrm C} - \frac{4}{5} V^{\pi}_{\mathrm T}
\end{array}
\right) \, ,
\label{eq:pi_pot_2-+} 
\end{eqnarray} }

\pagebreak

The $\rho$ and $\omega$ potentials are 
{\scriptsize
\begin{align}
V^{v}_{0(0^-)} &= 
\left(
2V^{v}_{\mathrm C}-2V^{v}_{\mathrm T} +V^{v \prime}_{\mathrm C}
\right) , \\
V^{v}_{0(1^+)} &=
\left(
\begin{array}{cccc}
  -2V^{v}_{\mathrm C} +V^{v \prime}_{\mathrm C} & -\sqrt{2}
   V^{v}_{\mathrm T} & 4V^{v}_{\mathrm C} & -\sqrt{2} V^{v}_{\mathrm T}   \\
  -\sqrt{2} V^{v}_{\mathrm T} & -2V^{v}_{\mathrm C}+V^{v}_{\mathrm T}+V^{v \prime}_{\mathrm C}  & -\sqrt{2} V^{v}_{\mathrm T} & 4V^{v}_{\mathrm C} + V^{v}_{\mathrm T} \\
 4V^{v}_{\mathrm C} & -\sqrt{2} V^{v}_{\mathrm T}  & 
-2V^{v}_{\mathrm C} +V^{v \prime}_{\mathrm C} & -\sqrt{2} V^{v}_{\mathrm T} \\
 -\sqrt{2} V^{v}_{\mathrm T} & 4V^{v}_{\mathrm C} + V^{v}_{\mathrm T} &
  -\sqrt{2} V^{v}_{\mathrm T} & -2V^{v}_{\mathrm C} + V^{v}_{\mathrm T}
  +V^{v \prime}_{\mathrm C} 
\end{array}
\right) , \\
V^{v}_{0(1^-)} &=
\left(
\begin{array}{ccccc}
  V^{v \prime}_{\mathrm C} & 0 & -2\sqrt{3} V^{v}_{\mathrm C} &
   2\sqrt{\frac{3}{5}}  V^{v}_{\mathrm T} & -3\sqrt{\frac{2}{5}}
   V^{v}_{\mathrm  T}   \\
  0 & 2V^{v}_{\mathrm C}+V^{v}_{\mathrm T} +V^{v \prime}_{\mathrm C} &
   0  & 3 \sqrt{\frac{3}{5}}
   V^{v}_{\mathrm T} &  -3\sqrt{\frac{2}{5}} V^{v}_{\mathrm T} \\
 -2\sqrt{3} V^{v}_{\mathrm C} & 0 & -4 V^{v}_{\mathrm C} +V^{v \prime}_{\mathrm C} &
 -\frac{2}{\sqrt{5}}V^{v }_{\mathrm T} & \sqrt{\frac{6}{5}} V^{v}_{\mathrm T} \\
  2\sqrt{\frac{3}{5}} V^{v}_{\mathrm T} & 3\sqrt{\frac{3}{5}}
   V^{v}_{\mathrm T} &  -\frac{2}{\sqrt{5}}V^{v}_{\mathrm T} &
   2V^{v}_{\mathrm C} + \frac{7}{5} V^{v}_{\mathrm T} +V^{v
   \prime}_{\mathrm C} & -\frac{\sqrt{6}}{5} V^{v}_{\mathrm T} \\
  -3\sqrt{\frac{2}{5}} V^{v}_{\mathrm T} & -3\sqrt{\frac{2}{5}}
   V^{v}_{\mathrm T} & \sqrt{\frac{6}{5}} V^{v}_{\mathrm T} &
   -\frac{\sqrt{6}}{5} V^{v}_{\mathrm T} & 2V^{v}_{\mathrm C} +
   \frac{8}{5} V^{v}_{\mathrm T} +V^{v \prime}_{\mathrm C}
\end{array}
\right) , \\
V^{v}_{0(2^+)} &=
\left(
\begin{array}{cc}
  -2V^{v}_{\mathrm C}-V^{v}_{\mathrm T} +V^{v \prime}_{\mathrm C}& 4V^{v}_{\mathrm C}-V^{v}_{\mathrm T}    \\
  4V^{v}_{\mathrm C}-V^{v}_{\mathrm T} & -2V^{v}_{\mathrm
   C}-V^{v}_{\mathrm T} + V^{v \prime}_{\mathrm C}
\end{array}
\right) , \\
V^{v}_{0(2^-)}  &=
\left(
\begin{array}{cccc}
 2V^{v}_{\mathrm C} - \frac{1}{5} V^{v}_{\mathrm T} + V^{v
  \prime}_{\mathrm C}  & \frac{3\sqrt{6}}{5} V^{v}_{\mathrm T} &
  -\frac{3\sqrt{3}}{5} V^{v}_{\mathrm T} & \frac{6\sqrt{3}}{5} V^{v}_{\mathrm T}   \\
 \frac{3\sqrt{6}}{5} V^{v}_{\mathrm T} & 2V^{v}_{\mathrm
  C}-\frac{4}{5}V^{v}_{\mathrm T} + V^{v \prime}_{\mathrm C} & -\frac{3\sqrt{2}}{5} V^{v}_{\mathrm T} & \frac{6\sqrt{2}}{5} V^{v}_{\mathrm T} \\
 -\frac{3\sqrt{3}}{5} V^{v}_{\mathrm T} & -\frac{3\sqrt{2}}{5}
  V^{v}_{\mathrm T}  &  2V^{v}_{\mathrm C} - \frac{7}{5} V^{v}_{\mathrm
  T} + V^{v \prime}_{\mathrm C} & -\frac{6}{5} V^{v}_{\mathrm T} \\
 \frac{6\sqrt{3}}{5} V^{v}_{\mathrm T} & \frac{6\sqrt{2}}{5}
  V^{v}_{\mathrm T} & -\frac{6}{5} V^{v}_{\mathrm T} & 2V^{v}_{\mathrm
  C} + \frac{2}{5} V^{v}_{\mathrm T} + V^{v \prime}_{\mathrm C}
\end{array}
\right) , \\
V^{v}_{1(0^+)} &=
\left(
\begin{array}{ccc}
  V^{v \prime}_{\mathrm C} & -2\sqrt{3} V^{v}_{\mathrm C}  & -\sqrt{6} V^{v}_{\mathrm T}   \\
  -2\sqrt{3} V^{v}_{\mathrm C} &- 4V^{v}_{\mathrm C} +V^{v \prime}_{\mathrm C} & \sqrt{2} V^{v}_{\mathrm T}  \\
 -\sqrt{6} V^{v}_{\mathrm T} & \sqrt{2} V^{v}_{\mathrm T}  
&  2V^{v}_{\mathrm C} +2V^{v}_{\mathrm T} +V^{v \prime}_{\mathrm C}
\end{array}
\right), \\
V^{v}_{1(0^-)} &=
\left(
\begin{array}{cc}
  -2V^{v}_{\mathrm C}+2V^{v}_{\mathrm T} +V^{v \prime}_{\mathrm C} & 4V^{v}_{\mathrm C}+2V^{v}_{\mathrm T}    \\
  4V^{v}_{\mathrm C}+2V^{v}_{\mathrm T} & -2V^{v}_{\mathrm
   C}+2V^{v}_{\mathrm T} +V^{v \prime}_{\mathrm C}
\end{array}
\right) , \\
V^{v}_{1(1^+)} &=
\left(
\begin{array}{ccc}
  2V^{v}_{\mathrm C}+V^{v \prime}_{\mathrm C} & \sqrt{2}
   V^{v}_{\mathrm T} & \sqrt{6}V^{v}_{\mathrm T} \\
  \sqrt{2} V^{v}_{\mathrm T} & 2V^{v}_{\mathrm C} -V^{v}_{\mathrm T}  +V^{v \prime}_{\mathrm C}&
  \sqrt{3} V^{v}_{\mathrm T}  \\
 \sqrt{6} V^{v}_{\mathrm T} & \sqrt{3} V^{v}_{\mathrm T} &
  2V^{v}_{\mathrm C}+ V^{v}_{\mathrm T} +V^{v \prime}_{\mathrm C}
\end{array}
\right) , \\
V^{v}_{1(1^-)} &=
\left(
\begin{array}{cc}
  -2V^{v}_{\mathrm C}-V^{v}_{\mathrm T} +V^{v \prime}_{\mathrm C}& -4V^{v}_{\mathrm C}+V^{v}_{\mathrm T}    \\
  -4V^{v}_{\mathrm C}+V^{v}_{\mathrm T} & -2V^{v}_{\mathrm
   C}-V^{v}_{\mathrm T} + V^{v \prime}_{\mathrm C}
\end{array}
\right) , \\
V^{v}_{1(2^+)}  & = 
 \left( 
\begin{array}{cccccc}
  V^{v \prime}_{\mathrm C} & 0 & -2\sqrt{3} V^{v}_{\mathrm C} &
   -\sqrt{\frac{6}{5}} V^{v}_{\mathrm T} & 2\sqrt{\frac{3}{7}}
   V^{v}_{\mathrm T} & -6\sqrt{\frac{3}{35}} V^{v}_{\mathrm T} \\
  0 & 2V^{v}_{\mathrm C}+V^{v}_{\mathrm T} +V^{v \prime}_{\mathrm C}& 0 & -3\sqrt{\frac{2}{5}}
   V^{v}_{\mathrm T} & \frac{3}{\sqrt{7}} V^{v}_{\mathrm T} &  \frac{12}{\sqrt{35}} V^{v}_{\mathrm T} \\
  -2\sqrt{3} V^{v}_{\mathrm C} & 0 & -4 V^{v}_{\mathrm C} +V^{v
   \prime}_{\mathrm C} &
   \sqrt{\frac{2}{5}} V^{v}_{\mathrm T} & -\frac{2}{\sqrt{7}}
   V^{v}_{\mathrm T} &  \frac{6}{\sqrt{35}} V^{v}_{\mathrm T} \\
  -\sqrt{\frac{6}{5}} V^{v}_{\mathrm T} & 3\sqrt{\frac{3}{5}}
   V^{v}_{\mathrm T} &  \sqrt{\frac{2}{5}} V^{v}_{\mathrm T} &
   2V^{v}_{\mathrm C} +V^{v \prime}_{\mathrm C} & -\sqrt{\frac{14}{5}} V^{v}_{\mathrm T} & 0 \\
  2\sqrt{\frac{3}{7}} V^{v}_{\mathrm T} & \frac{3}{\sqrt{7}}
   V^{v}_{\mathrm T} & -\frac{2}{\sqrt{7}} V^{v}_{\mathrm T} &
   -\sqrt{\frac{14}{5}} V^{v}_{\mathrm T} & 2V^{v}_{\mathrm C} -
   \frac{3}{7} V^{v}_{\mathrm T} +V^{v \prime}_{\mathrm C} & -\frac{12}{7\sqrt{5}} V^{v}_{\mathrm T} \\
  -6\sqrt{\frac{3}{35}} V^{v}_{\mathrm T} & \frac{12}{\sqrt{35}}
   V^{v}_{\mathrm T}  & \frac{6}{\sqrt{35}} V^{v}_{\mathrm T} & 0 &
   -\frac{12}{7\sqrt{5}} V^{v}_{\mathrm T} & 2 V^{v}_{\mathrm C} +
   \frac{10}{7} V^{v}_{\mathrm T} +V^{v \prime}_{\mathrm C}
\end{array}
\right) , \\
V^{v}_{1(2^-)} &=
\left(
\begin{array}{cccc}
  -2V^{v}_{\mathrm C} + \frac{1}{5} V^{v}_{\mathrm T} +V^{v
   \prime}_{\mathrm C} & -\frac{3\sqrt{6}}{5} V^{v}_{\mathrm T} &
   4V^{v}_{\mathrm C} + \frac{1}{5}V^{v}_{\mathrm T} & -\frac{3\sqrt{6}}{5} V^{v}_{\mathrm T}   \\
  -\frac{3\sqrt{6}}{5} V^{v}_{\mathrm T} & -2V^{v}_{\mathrm C}
   +\frac{4}{5}V^{v}_{\mathrm T} +V^{v \prime}_{\mathrm C}  & -\frac{3\sqrt{6}}{5} V^{v}_{\mathrm T} & 4V^{v}_{\mathrm C} + \frac{4}{5} V^{v}_{\mathrm T} \\
 4V^{v}_{\mathrm C} + \frac{1}{5} V^{v}_{\mathrm T} &
  -\frac{3\sqrt{6}}{5} V^{v}_{\mathrm T}  & -2V^{v}_{\mathrm C} +
  \frac{1}{5} V^{v}_{\mathrm T} + V^{v \prime}_{\mathrm C} & -\frac{3\sqrt{6}}{5} V^{v}_{\mathrm T} \\
 -\frac{3\sqrt{6}}{5} V^{v}_{\mathrm T} & 4V^{v}_{\mathrm C} +
  \frac{4}{5} V^{v}_{\mathrm T} & -\frac{3\sqrt{6}}{5} V^{v}_{\mathrm T}
  & -2V^{v}_{\mathrm C} + \frac{4}{5} V^{v}_{\mathrm T} + V^{v \prime}_{\mathrm C}
\end{array}
\right) .
\end{align}}
where the central and tensor potentials  are defined as,
\begin{eqnarray}
V^{\pi}_{\mathrm C} &=& \left( \sqrt{2}\frac{g}{f_{\pi}}\right)^2
\frac{1}{3}C(r;m_{\pi})  \vec{\tau}_{1} \!\cdot\! \vec{\tau}_{2} \, , 
\label{eq:Vcpi}\\
V^{\pi}_{\mathrm T} &=& \left( \sqrt{2}\frac{g}{f_{\pi}}\right)^2
\frac{1}{3}T(r;m_{\pi})  \vec{\tau}_{1} \!\cdot\! \vec{\tau}_{2} \, , \\
V^{\rho}_{\mathrm C} &=& \left( 2\lambda g_V \right)^2 \frac{1}{3}
C(r;m_{\rho}) \vec{\tau}_{1} \!\cdot\! \vec{\tau}_{2} \, , \\
V^{\omega}_{\mathrm C} &=&
 \left( 2\lambda g_V \right)^2 \frac{1}{3}C(r;m_{\omega}) \, ,  \\
V^{\rho}_{\mathrm T} &=& \left( 2\lambda g_V \right)^2 \frac{1}{3}
T(r;m_{\rho}) \vec{\tau}_{1} \!\cdot\! \vec{\tau}_{2} \, , \\
V^{\omega}_{\mathrm T} &=&
 \left( 2\lambda g_V \right)^2 \frac{1}{3}T(r;m_{\omega}) \, ,  \\
 V^{\rho \prime}_{\mathrm C} &=& \left( \frac{\beta g_V}{2m_{\rho}} \right)^2
\frac{1}{3} C(r;m_{\rho})\vec{\tau}_{1} \!\cdot\! \vec{\tau}_{2} \, , \\
V^{\omega \prime}_{\mathrm C} &=&
 \left( \frac{\beta g_V}{2m_{\omega}} \right)^2
\frac{1}{3} C(r;m_{\omega}) \, \label{eq:Vc'omega}. 
\end{eqnarray}

\end{document}